\documentclass[acmsmall,screen]{acmart}
\usepackage{subcaption}
\usepackage{tabularx}

\AtBeginDocument{%
  \providecommand\BibTeX{{%
    \normalfont B\kern-0.5em{\scshape i\kern-0.25em b}\kern-0.8em\TeX}}}

\acmPrice{15.00}
\acmISBN{978-1-4503-XXXX-X/18/06}

\usepackage{soul}
\usepackage{xspace}

\newcolumntype{R}{>{\raggedleft\arraybackslash}X}
\newcolumntype{L}{>{\raggedright\arraybackslash}X}
\usepackage{xcolor,colortbl}

\definecolor{Qanon}{rgb}{223,137,255}
\definecolor{Gray}{gray}{0.85}

\newcolumntype{q}{>{\columncolor{Gray}}c}

\begin{document}
\setcopyright{acmlicensed}
\acmJournal{PACMHCI}
\acmYear{2023} \acmVolume{7} \acmNumber{CSCW2} \acmArticle{252} \acmMonth{10} \acmPrice{15.00}\acmDOI{10.1145/3610043}
\title[A Golden Age]{A Golden Age: Conspiracy Theories' Relationship with Misinformation Outlets, News Media, and the Wider Internet}


\author{Hans W. A. Hanley}
\email{hhanley@stanford.edu}
\affiliation{%
  \institution{Computer Science Department, Stanford University}
  \streetaddress{450 Serra Mall}
  \city{Stanford}
  \state{California}
  \country{USA}
  \postcode{94305}
}
\author{Deepak Kumar}
\email{kumarde@stanford.edu}
\affiliation{%
  \institution{Computer Science Department, Stanford University}
  \streetaddress{450 Serra Mall}
  \city{Stanford}
  \state{California}
  \country{USA}
  \postcode{94305}
}

\author{Zakir Durumeric}
\email{zakird@stanford.edu}
\affiliation{%
  \institution{Computer Science Department, Stanford University}
  \streetaddress{450 Serra Mall}
  \city{Stanford}
  \state{California}
  \country{USA}
  \postcode{94305}
}
\renewcommand{\shortauthors}{Hans W. A. Hanley, Deepak Kumar, \& Zakir Durumeric}

\received{July 2022}
\received[revised]{January 2023}
\received[accepted]{March 2023}
\begin{abstract}
Do we live in a ``Golden Age of Conspiracy Theories?'' In the last few decades, conspiracy theories have proliferated on the Internet with some having dangerous real-world consequences. A large contingent of those who participated in the January 6th attack on the US Capitol fervently believed in the QAnon conspiracy theory. In this work, we study the relationships amongst five prominent conspiracy theories (QAnon, COVID, UFO/Aliens, 9/11, and Flat-Earth) and each of their respective relationships to the news media, both authentic news and misinformation. Identifying and publishing a set of 755 different conspiracy theory websites dedicated to our five conspiracy theories, we find that each set often hyperlinks to the same external domains, with COVID and QAnon conspiracy theory websites having the largest amount of shared connections. Examining the role of news media, we further find that not only do outlets known for spreading misinformation hyperlink to our set of conspiracy theory websites more often than authentic news websites but also that this hyperlinking increased dramatically between 2018 and 2021, with the advent of QAnon and the start of COVID-19 pandemic. Using partial Granger-causality, we uncover several positive correlative relationships between the hyperlinks from misinformation websites and the popularity of conspiracy theory websites, suggesting the prominent role that misinformation news outlets play in popularizing many conspiracy theories.
\end{abstract}



\maketitle

\section{Introduction}

%
In the last decade, websites have spread non-evidence-based conspiratorial ideas in an unprecedented fashion, resulting in a supposed ``Golden Age of Conspiracy''~\cite{Rao2022,Stanton2020}. Several websites, originally known for touting merely misleading information, have increasingly utilized conspiracy theories for specific political purposes~\cite{Ovide2021,PRRI2021,infowars-conspiracy}. Supporting this idea, today, almost 90\% of the American public endorse at least one conspiracy theory~\cite{Enders_Uscinski_Klofstad_Seelig_Wuchty_Murthi_Premaratne_Funchion_2021,vermeule2009conspiracy,oliver2014medical}, including those about 9/11~\cite{stempel2007media}, UFOs/Aliens~\cite{dean1998aliens}, Flat-Earth~\cite{paolillo2018flat}, COVID-19~\cite{romer2020conspiracy}, and QAnon~\cite{amarasingam2020qanon}. According to a poll by the Public Religion Institute and the Interfaith Youth Core in July 2021, approximately 15\% of Americans are even true ``QAnon believers'' (\textit{i.e.}, those who believe that ``the government, media, and financial worlds in the U.S. are controlled by a group of Satan-worshipping pedophiles who additionally run a global child sex trafficking operation''~\cite{PRRI2021}). 

Researchers have found that individuals mostly gain exposure to conspiracy theories like QAnon through the Internet~\cite{sunstein2018republic,uscinski2014american}. Indeed, websites like the Gateway Pundit, American Thinker, and InfoWars have played an outsized role in developing toxic echo chambers~\cite{starbird2018ecosystem} 
 and may have also exposed thousands of people to different conspiracy theories. Prior research has shown that mere exposure to conspiracy theories can unconsciously influence people to believe in them~\cite{douglas2008hidden}. The exposure and subsequent belief in just one conspiracy theory can further induce paranoia and stimulate belief in other conspiracy theories~\cite{hofstadter2012paranoid, goertzel1994belief,bruckmuller2017past,bale2007political,zonis1994conspiracy}. However, despite conspiracy theories' increasing prevalence, the precise relationship each conspiracy theory has with one another and the relationship between the coverage of conspiracy theories by online news outlets and the popularity of conspiracy theories is largely unknown. However, in this work, we use web crawling, hyperlink graphs, and statistical models of websites' behavior to examine the relationships between  conspiracy theories and the role that online misinformation outlets, authentic news media, and the wider Internet ecosystems play in their coverage, asking the following three research questions:

\begin{enumerate}
    \item \textit{How do particular conspiracy theories hyperlink/interact with one another, the news media, and the wider Internet? }
    \item \textit{How has news media's (both authentic news platforms and misinformation outlets) relationship with particular conspiracy theories changed over time? Have different outlets' interactions with these conspiracies increased or decreased over time?}
    \item{\textit{Has the news media's (both authentic news platforms and misinformation outlets) interaction with particular conspiracy theories correlated with the rise or decline of specific conspiracy theories' popularity?}}
 
\end{enumerate}

To answer these three questions, we curate and analyze the behavior of a set of 755~conspiracy theory websites, subdivided into five different categories: QAnon (227~websites), COVID (134), UFO/Aliens (193),  9/11 (104), Flat-Earth (97). Using our own web scrapes and pages scraped by Common Crawl,\footnote{\url{https://commoncrawl.org/}} we then document the state and the changing behaviors of the conspiracy theory ecosystem and their relationship to a separate set of 530~known misinformation outlets, 565~authentic news websites, and 528~non-news websites.

\textbf{RQ1: The State of the Conspiracy Theory Ecosystem and Its Relationship with News Media:} To understand how different conspiracy theories interact amongst themselves, we first examine the shared hyperlink connections that our conspiracy theory categories have with each other. Across all the different conspiracy theory domains considered, we find that QAnon-focused websites have the highest percentage of shared connections with other conspiracy theory website groups, particularly COVID (35.6\%); this largely accords with news reporting that QAnon has become a ``big tent conspiracy theory'' that incorporates the beliefs of other conspiracy theories~\cite{Roose2021}.  Utilizing the Common Crawl harmonic and PageRank centrality measures that measure websites' centrality across all of the \emph{crawled} Internet, we then find many of the conspiracy theory-focused websites in our dataset have relatively high network centrality, suggesting that many of them are not peripheral on the Internet but actually near the Internet's core/are mainstream. Indeed examining, the hyperlink connections between news media and these conspiracy theories, we find that many of them rely heavily on authentic news as well as misinformation outlets (compared to non-news websites) for their information, with many popular misinformation outlets also hyperlinking back to many of these conspiracy theory websites.

\textbf{RQ2: News Media's Changing Relationship With Conspiracy Theories} Having observed the relatively strong relationship between our set of conspiracy theory websites and different news outlets, we next seek to understand whether these connections have changed over time. We find, starting in 2018, a significant increase in hyperlinks to our set of conspiracy theory websites with the advent of QAnon and then again in 2019-2020 with the beginning of the COVID-19 pandemic. Concurrent with this large increase in hyperlinks to our set of websites we further observe similar increases in the popularity of our set of conspiracy theory websites, and a \textbf{more general increase from misinformation outlets towards \textit{conspiracy-oriented} material}. Indeed, examining an additional set of 116K fringe websites, we find that between 2009 and 2021, the percentage of all misinformation websites' external hyperlinks to these \textit{conspiracy-oriented} websites went from 9.0\% to 13.2\%, a 46.6\% relative increase.

\textbf{RQ3: Misinformation Outlets' Role in Promoting Conspiracy Theories:} We finally apply partial Granger causality analysis to ascertain whether the behavior of misinformation websites is a factor in the popularity (\textit{i.e.}, Amazon Alexa Rank~\cite{amazon-top-mil}) of conspiracy theories online. Partial Granger causality is a means of measuring if a given time series is useful for forecasting another while taking into account unmeasured endogenous and exogenous factors~\cite{guo2008partial}. Our results suggest that in several cases as misinformation sites hyperlinked to conspiracy theory sites, this, in turn, correlated with the increased popularity of conspiracy theory websites. Conversely, we find, for QAnon in particular,  that as more mainstream outlets wrote about this conspiracy theory, the popularity of the corresponding QAnon-focused websites decreased in popularity.

Conspiracy theories have come to play an increasingly large role in world events, as was apparent in the partly QAnon-inspired January 6, 2021, attack on the U.S. capitol~\cite{Ovide2021}. Similarly, as COVID-19 spread throughout the world, conspiracy theories became a major impediment to curbing the pandemic~\cite{romer2020conspiracy}. Our analysis shows the role news outlets have had in promoting and spreading conspiracy theories throughout the past decade. As conspiracy theories continue to play a larger role on the Internet, we hope that our results help shed light on their spread and show the utility of online structure-based network analysis in understanding their ecosystems.




\section{Background and Related Work}\label{sec:background}

In this section, we give an overview of how we operationalize our study of authentic news platforms, misinformation outlets, conspiracy theories, and non-news websites. We further give background on previous studies that have studied similar ecosystems.

\subsection{Terminology and Operalization}
We first describe key definitions and concepts that we utilize throughout this work.  

\textbf{\textit{Domain Based Identification:}} As described by Abdali \textit{et al.}, examining misinformation, conspiracies, and news from the website domain level rather than by article level is a more ``fruitful'' enterprise~\cite{abdali2021identifying}.  The consistent publication of misinformation, conspiracy theories, and truthful information on a particular domain is more readily identifiable and can be examined over time~\cite{haninfrastructure,hanley2021no,hounsel2020identifying}. For example, it is unlikely that reputable organizations like The New York Times or the Wall Street Journal will start regularly publishing false information in the near future. In contrast, websites like The American Thinker or The Conservative Treehouse regularly and consistently publish false information. We, therefore, in this work, categorize and label websites on a domain level. 

\textbf{\textit{Misinformation Websites:}} False and misleading information or ``\textit{misinformation}'' takes a host of different forms including unintentional misreporting, deliberate hoaxes, pranks, political propaganda, and disinformation. Following the 2016 US Presidential election, Jack \textit{et al.}~\cite{jack2017lexicon}\ presented an in-depth categorization of different types of misinformation. Since this report, the two distinguishing features of different types of false information that have gained widespread acceptance are \textit{veracity} and \textit{intentionally}~\cite{10.1145/3137597.3137600, jiang2018linguistic}. Many works, using these criteria, define \textit{misinformation} as any information that is false or inaccurate regardless of the intention of the author~\cite{jiang2018linguistic,guess2018selective,huang2015connected,lewandowsky2012misinformation,weeks2015emotions,haninfrastructure,hanley2021no,allcott2019trends}. \textit{Disinformation}, meanwhile, is false and inaccurate information spread with the express and deliberate purpose to mislead. Similar to disinformation, \textit{propaganda} refers to ``deliberate, systematic information campaigns, usually conducted through mass media forms'' regardless of whether the information is true or false ~\cite{jack2017lexicon}.

We, as in previous works~\cite{haninfrastructure,hanley2021no,zannettou2017web,hounsel2020identifying,sharma2022construction,abdali2021identifying,paraschiv2022unified,cheng2021causal,allcott2019trends}, define \textit{misinformation websites} as news websites that regularly publish false information about current events, engage is propagandist campaigns meant to mislead, and that do not engage in journalistic norms such as attributing authors and correcting errors. As in these past works, we specifically \textit{choose} to utilize a definition of \textit{misinformation websites} that centers around misinformation, propaganda, \textit{and} disinformation about current events~\cite{hounsel2020identifying,Kasprak2019,Evon2019,allcott2019trends}. We do this to ascertain the relationship between prominent websites of this encompassing type that regularly publish false information and conspiracy theories. 

\textbf{\textit{Conspiracy Theory Websites:}}   A conspiracy theory is ``an explanation [of current events] that refers to hidden malevolent forces that seek to advance some nefarious aim''~\cite{Oliver2014}. Conspiracy theories have been often thought of as a subset of misinformation and disinformation, and we largely consider them as misinformation here as well. However, as outlined by others, most markedly by Brotherton \textit{et al.}, conspiracy theories have specific and distinct characteristics that make them different from general misinformation~\cite{brotherton2013measuring,butter2020routledge,douglas2019understanding} and thus are a particular aspect of misinformation. Specifically, Brotherton \textit{et al.}~\cite{brotherton2013measuring}, using factor analysis, found that the five key features of modern conspiracy theories are \textit{(1) government malfeasance, (2) extraterrestrial cover-up, (3) malevolent global conspiracies, (4) personal well-being, and (5) control of information.}  As found by Samory and Mitra~\cite{samory2018government}, conspiracy theories often also follow similar \textit{narrative-motifs} that focus on the conspiratorial \textit{agents}, the \textit{actions} these \textit{agents} perform, and the \textit{targets} or \emph{victims} of these actions. Due to this, among other particularities, it has been found that measures taken at correcting and stemming misinformation remain mostly ineffective against conspiracy theories~\cite{nyhan2010corrections,stojanov2015reducing,douglas2019understanding,bessi2014social}. 

To make our conspiracy theory category concrete, we populate this category with websites that are focused on or dedicated to five different conspiracy theories (QAnon, COVID, 9/11, UFO/Aliens, Flat-Earth) that fall under psychologist Bortherton's \textit{et al.}'s definition of conspiracy theories~\cite{brotherton2013measuring}. For this study, in order to examine how different and more general websites utilize specific conspiracy theories, we thus consider a website as \textit{misinformation} if it generally regularly publishes false information about current events (without it having a focus on any given conspiracy theory) while considering a website as \textit{conspiracy theory} website if it focuses on our five of our listed conspiracy theories. For instance, the vast majority of the 755~conspiracy theory websites in our data to be outlined later, directly contain the name of their conspiracy theory in their domain name. Establishing this distinction enables us to identify the role that specific conspiracy theories have in the spread of political news and misinformation as well as to operationalize the study of individual conspiracy theories on the Internet. In this way, as in prior works, we examine more general misinformation and news outlets' relationship with specific websites that push a given conspiracy theory~\cite{kong2022slipping,hanley2021no,papasavva2022gospel,sharma2022characterizing,phadke2022pathways,samory2018conspiracies,samory2018government,phadke2021characterizing}.

\textbf{\textit{Authentic-News:}} 
As in previous works, we define authentic news websites as outlets that generally adhere to journalistic norms including attributing authors and correcting errors; altogether publishing mostly true information ~\cite{hounsel2020identifying,hanley2021no,zannettou2017web}.

\textbf{\textit{Non-News:}} 
As in Hounsel \textit{et al.}~\cite{hounsel2020identifying}, we define \textit{non-news websites} as those that do not normally traffic in current-event-related topics.

\subsection{Related Work: The Impact of Misinformation}
Our work, which examines and measures the impact of specific social phenomena (prominent conspiracy theories and misinformation), builds on others' contributions that seek to identify the long-term influence of toxic social contagions. Several works have attempted to identify and understand the impact of misinformation utilizing metadata gleaned from microblogging messages~\cite{jain2016towards,hanley2022special,qazvinian2011rumor, ghenai2017catching}, network infrastructure~\cite{haninfrastructure}, natural language processing~\cite{serrano2020nlp, oshikawa2020survey, gilda2017notice}, and images~\cite{garimella2020images, zannettou2020characterizing, wang2020understanding}. For example, Zannettou \textit{et~al.}\ and Wang \textit{et~al.}\ studied misinformation through shared images on social media ~\cite{zannettou2020characterizing,wang2020understanding}. Taking a different image-based approach, Abdali \textit{et~al.}\ identified misinformation websites from screenshots of their articles~\cite{abdali2021identifying}. Qazvinian \textit{et al.}\ helped pioneer approaches utilizing text-based linguistic features by identifying particular phrases present in misinformation articles~\cite{qazvinian2011rumor}. In contrast to these previous approaches, several others have taken a more case-study-based approach to studying the influence of particular misinformation campaigns. For example, Wilson and Starbird \textit{et al.} look at the spread of disinformation campaigns related to the Syrian White Helmets~\cite{starbird2018ecosystem}.


Like our work, many works have exploited the semantic information that exists in mutual hyperlinking amongst semantically similar domains to study online behaviors and misinformation. As shown in Bhatt \textit{et al.}, Hanley \textit{et al.},  Garimella \textit{et al.}, and Miller \textit{et al.}, semantic information is often embedded within hyperlink graphs ~\cite{bhatt2018illuminating,hanley2021no,garimella2021political,miller2011sentiment}. Hanley \textit{et al.} showed that QAnon-themed websites often had deeper connections to misinformation compared to authentic news. Garimella \textit{et al.} showed a 0.70~correlation between a given website's polarization level and the average polarization levels of websites linked to it. Sehgal \textit{et~al.}\ show that hyperlinks shared between websites and Twitter users can be utilized to identify coordinated efforts to spread disinformation~\cite{sehgal2021mutual}. Their work buttresses these results by building a classifier based on mutual link sharing amongst misinformation spreaders on Twitter.

\subsection{Related Work: The Spread of Conspiracy Theories}
This work seeks to understand changes in the prevalence of five different conspiracy theories online during the past decade. While research literature indicates that the predilection of people to believe in conspiracy theories has remained somewhat constant across time~\cite{uscinski2014american,van2014power}, various authors have concluded that large spikes in conspiratorial belief often occur following major societal change~\cite{hofstadter2012paranoid,van2017conspiracy}. 
 As in our work, Bessi \textit{et~al.}~\cite{bessi2015science} examine the insular communities and polarized communities that come together to discuss and engage with news related to conspiracy theories. Samory and Mitra~\cite{samory2018conspiracies} use causal time-series analysis to understand changes in conversation and engagement following dramatic world events on the Reddit sub-community \texttt{r/conspiracy}. They find that after these dramatic events, users on Reddit's \texttt{r/conspiracy} subcommunity show signs of emotional shock as well as increased expressions of both certainty and doubtfulness.
 
 While many websites online are known to promote conspiracy theories~\cite{xiao2021sensemaking,uscinski2014american,sunstein2018republic}, these beliefs are not new.  In the United States, conspiracy theories from those about the death of John F. Kennedy to the Masonry to Watergate have possessed the American mind. According to Gallup, at a high point in the early 2000s, approximately 81\% of the American public believed in a conspiracy theory about the JFK assassination ~\cite{swift2013majority}.

\section{Dataset}
\label{sec:dataset}
Having given background and outlined the terminology used throughout this paper, we now describe the sets of website lists and data we collect utilizing the definitions previously provided. Our website lists are independent and non-overlapping.

\begin{table}
    \small
    \centering
    \begin{tabularx}{.7\columnwidth}{Xlr}
    \toprule
    {Conspiracy Theory} & {Seed Websites} & \# Websites  \\
    \midrule
    {QAnon}      &  {8kun.top, voat.co} & 227\\
    {UFO/Aliens} & {ufoabduction.com} & 193 \\
    {COVID}  &  {covid-is-fake.blogspot.com} & 134\\
    {9/11 }& {911truth.org} &  104 \\
    \mbox{Flat-Earth} & {theflatearthsociety.org} &  97 \\
    \bottomrule
    \end{tabularx}
   \caption{\textbf{Collecting conspiracy theories}--- Using Google searches, we collect a set of seed websites before performing deep crawling and graph analysis to later manually identify conspiracy theory websites in each category.}
   \label{table:conspiracy-collection}
   \vspace{-10pt}

\end{table}

\subsection{Conspiracy Theory Websites}\label{sec:conspiracy-focused-dataset}

As previously stated, a conspiracy theory is ``an explanation [of current events] that refers to hidden malevolent forces that seek to advance some nefarious aim''~\cite{Oliver2014}. Our study focuses on five conspiracy theories whose websites are readily identifiable:

\begin{enumerate}
    \item \textbf{QAnon}: 
    This conspiracy holds that the U.S. government is run by a cabal of Satanic pedophiles~\cite{amarasingam2020qanon}. This conspiracy theory falls under Brotherton \textit{et~al.'s} conspiracy features of government malfeasance, global conspiracies, and control of information~\cite{brotherton2013measuring}.
    \item \textbf{COVID}: A set of theories that promote fake cures for COVID-19, argue that COVID-19 is a fake illness, or spread the idea that COVID-19 vaccines are deadly poisons~\cite{romer2020conspiracy}. This conspiracy theory falls under Brotherton \textit{et al.'s} conspiracy features of government malfeasance, malevolent global conspiracies, personal well-being, and control of information.
    \item \textbf{9/11}: A conspiracy theory that holds that the 9/11 terrorist attack was a false flag event planned and orchestrated by the U.S. government or a world government~\cite{stempel2007media}. This conspiracy theory falls under Brotherton \textit{et~al.'s} conspiracy features of government malfeasance, malevolent global conspiracies, and control of information.
    \item \textbf{UFO/Aliens}: A conspiracy theory that holds that the United States government is secretly hiding that intelligent alien life is real and that aliens have repeatedly visited Earth~\cite{dean1998aliens}. This conspiracy theory falls under Brotherton \textit{et~al.'s} conspiracy features of extraterrestrial cover-up,  malevolent global conspiracies, and control of information.
    \item \textbf{Flat Earth}: A conspiracy that promotes that the Earth is flat and not spherical~\cite{paolillo2018flat}. Images taken of the earth from space as well as other evidence of the Earth being round are part of an elaborate hoax orchestrated by the world's governments and scientists~\cite{Pappas2021}. This conspiracy theory falls under Brotherton \textit{et~al.'s }conspiracy features of malevolent global conspiracies and control of information.
\end{enumerate}

\noindent
To the best of our knowledge, while there exist lists that do label certain websites as promoting conspiracy theories,\footnote{\url{https://github.com/several27/FakeNewsCorpus}} there are no public datasets that provide granular information about which specific conspiracy theory (\textit{i.e.}, QAnon, 9/11) a given website promotes. We thus, to create lists of this specificity, rely on an approach outlined by Hanley \textit{et~al.}~\cite{hanley2021no}, which leverages seed sets of websites and web crawling to find semantically-related websites. Originally only done for QAnon in Hanley \textit{et~al.}~\cite{hanley2021no}, we leverage this methodology five separate times, once for each of our conspiracy theories.

To build seed sets of websites per Hanley \textit{et~al.}'s methodology, we rely on Google searches, identifying 1--2~websites for each of our conspiracy theories (Table~\ref{table:conspiracy-collection}). After crawling each seed site and the websites that they hyperlinked, we identified additional candidate conspiracy theory websites from the hyperlinked domains using graph-based overlap node similarity with the original seed websites~\cite{vijaymeena2016survey} (calculated using their set of shared domains connections with the seed websites). This metric helped to create a list of websites with semantically similar content to the original seed set. After identifying candidates, we manually reviewed each twice,  confirmed each candidate's website was truly related to its given conspiracy theory in each run of this algorithm, and subsequently removed \emph{all} websites about which there was disagreement.  While costly in terms of scraping time, this approach enabled us to build five separate lists of websites dedicated to each of our five conspiracy theories. We crawled, scraped, and built our lists of conspiracy websites throughout June 2021. We release all these websites at \url{https://github.com/hanshanley/Golden-Age}.

We finally note that we take a relatively conservative approach in labeling our conspiracy theory websites, utilizing the approach outlined in Dahlke \textit{et~al.}~\cite{dahlke2022mixed}. Namely, we label websites as conspiracy theory websites if they ``explicitly acknowledge
their relationship to [the conspiracy theory] by hosting discussions of [the conspiracy theory]~\cite{dahlke2022mixed}.'' As such, we only consider a website to fit in these categories if it is most prominently displayed or is centered around its given conspiracy theory. For example, while 8kun.top was useful for finding QAnon-related websites, it also has forums dedicated to COVID, Ebola, and UFOs, could not be cleanly categorized, and is not included within our set of 227 QAnon domains.

Given this criteria, we note that 101 of our 104 nine-eleven websites mention ``911'' in their domain name. 94 out of our set of 97 flat earth websites mention ``flat'' or ``flat-earth'' in their domain name. 118 of our 135 COVID domains mention ``corona'' or ``covid'' or ``vaccine'' in their name. All 193 UFO/Aliens websites contain utilize ``ufo'' or ``alien'' in their domain name. Lastly, given a great deal of scholarship surrounding QAnon websites, we cross-reference our list of QAnon websites with previously curated lists to ensure its accuracy. All QAnon websites were found within these previously created lists~\cite{hanley2021no,wang2022identifying,papasavva2021qoincidence,papasavva2022gospel}. In total, we collected 755~conspiracy theory websites: 227~QAnon, 134~COVID, 104~Nine-Eleven, 193~UFO/Aliens, 97~Flat-Earth.

\subsection{Misinformation News Dataset} 
Much investigation has gone into identifying misinformation websites, and we utilize several previously curated lists from prior research~\cite{hanley2021no,haninfrastructure,hounsel2020identifying}. Specifically, we select websites from lists created by Iffy News,\footnote{\url{https://iffy.news/index/}} OpenSources,\footnote{\url{https://github.com/several27/FakeNewsCorpus}} Politifact,\footnote{\url{https://www.politifact.com/article/2017/apr/20/politifacts-guide-fake-news-websites-and-what-they/}} Snopes,\footnote{\url{https://github.com/Aloisius/fake-news}} and Melissa Zimdars.\footnote{\url{https://library.athenstech.edu/fake}} We filter these lists to websites labeled as ``fake news'', ``fake'', ``unreliable'', or that have a factual rating of ``low'' and ``very-low.'' We group these categories together given our definition of misinformation as news websites that regularly publish false information (Section~\ref{sec:background}). Combining these filtered lists together, we manually verify that each of the \textit{misinformation} websites in the combined list do not focus or are dedicated to our set of conspiracy theories. Finally, after removing duplicates and inactive websites, we arrive at 530~misinformation websites. This list of misinformation websites includes websites that have been widely documented as promoting falsehoods including The American Thinker and The Conservative Treehouse~\cite{starbird2018ecosystem}.

\begin{table}
    \small
    \centering
    \begin{tabularx}{.7\columnwidth}{Xr}
    \toprule
    {Website Category} & Total Websites  \\
    \midrule
    {Misinformation News}  & 530\\
    {Authentic News} & 565 \\
    {Non-News}  &  528\\
    \bottomrule
    \end{tabularx}
  \caption{\textbf{Curating website lists}--- We utilize several previously curated lists including those of the Columbia Journalism Review, OpenSources, Politifact, Snopes, and Melissa Zidmars to create lists of misinformation news and authentic news websites. We utilize semanticweb.org to create a new list of non-news-focused websites.}
   \label{table:website-cllections}
   \vspace{-10pt}
\end{table}

\subsection{Authentic News Dataset}
For our list of authentic news websites, we again utilize lists curated by OpenSources, Politifact, Snopes, and Melissa Zimdars, combining the sets of websites labeled as ``reliable'' (\textit{e.g.}, theguardian.com, nytimes.com, \textit{etc...}) or simply ``biased'' (\textit{e.g.}, dailywire.com.)\footnote{We include ``biased'' websites as they are more often just hyperpartisan rather than misinformation}. To supplement our list of authentic news websites, we also include a selection of local news websites from lists of U.S. newspapers~\footnote{\url{https://www.w3newspapers.com/}}\footnote{\url{https://en.wikipedia.org/wiki/List\_of\_newspapers\_in\_the\_United\_States}} such as the Orlando Sentinel, the Denver Post, and the Tampa Bay Times. Finally, we expand our list by including the websites of 61~Pulitzer Prize-winning newspapers. Our final list of 565 websites includes sites across the political spectrum, from dailywire.com to cnn.com. 

\subsection{Non-News Dataset}
We utilize 528 popular non-news websites as a control group to isolate and understand how conspiracy theories interact with the wider Internet. We include a subset of websites from  11~different topic category lists as labeled and gathered by {semanticweb.com}: business and services, community and society, food and drink, health, heavy industry, lifestyle, pets and animals, science and education, sports, travel and tourism, and vehicles.




\section{Methodology}\label{sec:methodology}

Having described the root set of websites that we analyze, we now present our full methodology for gathering data about and understanding the relationships between these websites.


\subsection{Common Crawl Page Retrieval and Website Crawling}
To gather the set of hyperlinks between our websites, we utilize Common Crawl data~\cite{smith2013dirt}---widely considered the most complete publicly available source of web crawl data---and our own website crawls. For each website in our dataset, we collect \textit{all} the domain's HTML pages that were indexed by Common Crawl before August 2021. In addition to Common Crawl data, we further utilize our own website scrapes. We utilize our own crawls, in addition to Common Crawl, due to noisiness, missing pages, and missing domains within the Common Crawl dataset~\cite{seddah2020building}. For example, 309 particularly small conspiracy theory domains were not contained within the Common Crawl dataset (\textit{i.e.}, these websites often only contained a few dozen pages). Thus for each website in our dataset, we further gather all the HTML pages 10~hops from each website's homepage (\textit{i.e.}, we collect all URLs linked from the homepage (1st hop), then all URLs linked from the pages that were linked by the homepage (2nd hop), and so forth). For each HTML page from our scrapes and Common Crawl, we parse the HTML, detect the date that page was published with the Python library \texttt{htmldate}, and collect hyperlinks to other pages (\textit{i.e.}, HTML <a> tags). Altogether we gather the available Common Crawl pages and scrape the HTML for our 755~conspiracy theory, 530~misinformation, 565~authentic news, and 528~non-news websites.  We crawled, scraped, and built this dataset between June 2021 and October 2021.

\begin{figure}
\centering
\begin{minipage}[c]{0.5\textwidth}
\includegraphics[width=\columnwidth]{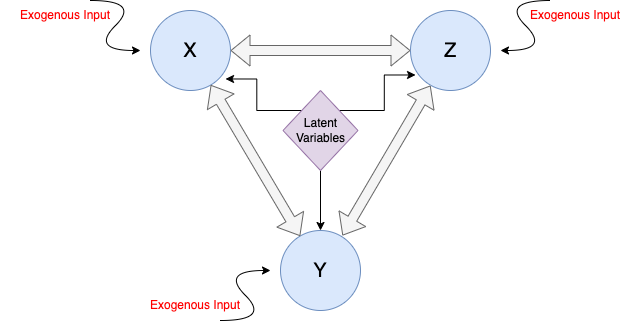}
\end{minipage}
\begin{minipage}[c]{0.45\textwidth}
\caption{\textbf{Modelling partial Granger causality relationships }--- Partial Granger causality models the interaction of multiple time series while taking into account exogenous environmental inputs and latent endogenous variables. }
\label{figure:partial-granger-ex}
\end{minipage}

\end{figure}
\subsection{Partial Granger Causality}\label{sec:methodology-granger}
Later in this work, to ascertain the relationship between authentic news, misinformation, and conspiracy theories, we perform a temporal causal analysis of how these three categories of websites interact. Specifically, after fitting our time-series data to linear vector autoregressive models (VAR), we utilize notions of partial Granger causality~\cite{granger1969investigating,chan1986estimating,guo2008partial} to determine if the behavior of websites in one of these groups is predictive of the behavior in another group. 

Granger causality is a \textbf{\textit{correlational}} means of measuring if one time series is useful for forecasting another. In practice, Granger causality is commonly tested utilizing linear vector autoregressive models (VAR) of two different time series $X_t$ and $Y_t$. VAR models forecast future values of a given time series based on its past values along with other variables including~\cite{chan1986estimating}. Using the VAR models, it is then tested if knowledge of time-series $Y_t$ increases the predictive power regarding $X_t$; if so, then $X$ Granger causes $Y$~\cite{granger1969investigating,guo2008partial}. For a more detailed description of Granger causality see Appendix~\ref{sec:granger-causality}.

Partial Granger causality is an extension of Granger causality for more than two time series that also takes into account latent endogenous variables and exogenous environmental variables~\cite{guo2008partial,yurdakul2015determinants}. More concretely, partial Granger causality models two time-series $X_t$ and $Y_t$ conditioned on a third time-series $Z_t$ while taking into account exogenous environmental inputs $\overrightarrow{\epsilon^E_t}$ as well as endogenous latent variables $\overrightarrow{\epsilon^L_t}$ of the three time-series. We picture this model in Figure~\ref{figure:partial-granger-ex}. Partial Granger causality then tests against the null hypothesis that the information present in $Y_t$ does not add information when predicting future values of $X_t$. 

Using this construct, for example, we can model the Granger-causal behavior of misinformation websites on the popularity of conspiracy theory websites while taking into account the behavior of authentic news websites as well as exogenous and potentially unmeasured and unaccounted-for endogenous variables. For a detailed overview of partial Granger causality see Appendix~\ref{sec:p-granger-causality}.

\subsection{Ethical Considerations}
Within this work, we largely look at large-scale trends amongst our set of websites and do not attempt to deanonymize website owners or particular article authors. We further utilize only public data and follow ethical guidelines for scraping websites as outlined by others~\cite{hanley2021no,acar2014web}.

\section{RQ1: The Conspiracy-Theory/News Ecosystem}\label{sec:ecosystem}

Having given an overview of our methodology, in this section, we now present an overview of the current state of the conspiracy theory ecosystem and the relationships between specific conspiracy theories. We then illustrate how different misinformation, authentic news, and non-news websites statically interact with these conspiracy theory websites. 

\subsection{The Conspiracy Theory Ecosystem}\label{sec:conspiracy-focused}
\begin{table}
    \small
    \centering
    \begin{tabularx}{.82\columnwidth}{l|l|l|l|l|l}
    \toprule
     & QAnon $\rightarrow$ &  Flat-Earth$\rightarrow$& COVID $\rightarrow$& 9/11 $\rightarrow$& UFO/Aliens $\rightarrow$ \\ \midrule
     QAnon $\leftarrow$ & -- & 7.5\%  & 14.9\% &  11.4\% & 12.2\% \\ 
     Flat-Earth $\leftarrow$& 19.1\% & -- & 11.2\% & 12.6\%& 15.4\%\\
     COVID $\leftarrow$ &  35.6\% & 10.3\% & -- &  16.1\%& 16.8\% \\
     9/11 $\leftarrow$  & 21.3\% &9.3\% & 13.2\% &  --- & 14.5\%\\
     UFOs/Aliens $\leftarrow$ & 15.5\% & 7.6\%  & 9.0\% & 9.7\% & -- \\
    \bottomrule
    \end{tabularx}
    \caption{\textbf{Percentage of website domains hyperlinked by each conspiracy theory category that are also hyperlinked by another category}--- Each conspiracy theory shares domain connections with each other. As seen, QAnon has the most shared domain connections with every other conspiracy theory.19.1\%, 35.6\%, 21.3\%, and 15.3\% of all domains hyperlinked by Flat-Earth, COVID, 9/11, and UFO/Aliens websites, respectively, are also hyperlinked by QAnon websites.}
\label{table:conspiracy-domain-connections}
\vspace{-10pt}
\end{table}

To determine the relationships between different conspiracy theories, in Table~\ref{table:conspiracy-domain-connections}, we list the percentage of outward domain connections that each conspiracy theory ecosystem shares with one another. This is such that we determine the percentage of domains hyperlinked by QAnon websites that are also hyperlinked by COVID conspiracy theory websites (\textit{e.g.}, 19.1\% of Flat-Earth's outward connections are also hyperlinked by QAnon websites).

\begin{figure}
\centering
\begin{subfigure}{.5\textwidth}
  \centering
  \includegraphics[width=\columnwidth]{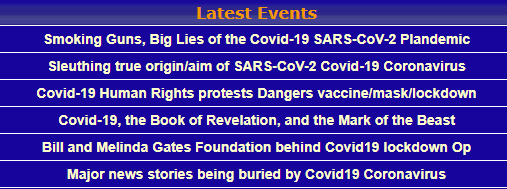}
\end{subfigure}
\caption{\textbf{Conspiracy Interaction: 9/11  to COVID}--- Example of how a 9/11 conspiracy theory website (\mbox{911forum.org.uk}) hyperlinks to and promotes COVID conspiracy theories.}
\label{figure:nine11-covid}
\vspace{-10pt}
\end{figure}
\textbf{\textit{Interdependence Among Conspiracy Theories:}} As seen in Table~\ref{table:conspiracy-domain-connections}, all of our conspiracy theories website groups, despite being independent, all share significant connections among themselves; every category of websites shares at least 7.5\% (at a minimum) of their connections with every conspiracy theory website group. This reinforces prior research that has suggested that people who believe one conspiracy theory are more likely to believe in others~\cite{goertzel1994belief}. Directly evidencing this phenomenon, as an example, the 9/11 conspiracy theory website 911forum.org.uk, has a direct reference to the COVID conspiracy “Plandemic” being referenced even on a 9/11 conspiracy theory website (Figure~\ref{figure:nine11-covid}).

\textbf{\textit{Strength of QAnon:}} Looking at each conspiracy theory's connections, we observe extensive interconnections between the QAnon and COVID conspiracy theories. 35.6\% (Table~\ref{table:conspiracy-domain-connections}) of all domains hyperlinked by COVID conspiracy theory websites are also hyperlinked by QAnon websites. As reported by The Washington Post and others, following the defeat of Donald Trump in the 2020 election, many QAnon groups began to focus their ire on COVID vaccines and the government lockdowns\cite{Timberg2021,morelock2022nexus}. This is largely matched by the elevated levels of interconnectivity between these two groups. We further see an elevated connection with QAnon across the rest of the conspiracy theories considered in this work. 21.3\%, 19.1\%, and 15.3\% of all domains hyperlinked by 9/11, Flat-Earth, and UFO/Aliens websites, respectively, are also hyperlinked by QAnon websites; each of them has the most shared connections with QAnon websites as opposed to the other conspiracy theories. This also largely accords with QAnon being a highly active conspiracy theory and news reporting that QAnon has become a ``big tent conspiracy theory'' that often incorporates the claims of other conspiracy theories~\cite{Roose2021}.

\begin{figure*}
\begin{subfigure}{.30\textwidth}
  \centering
  \includegraphics[width=1\linewidth]{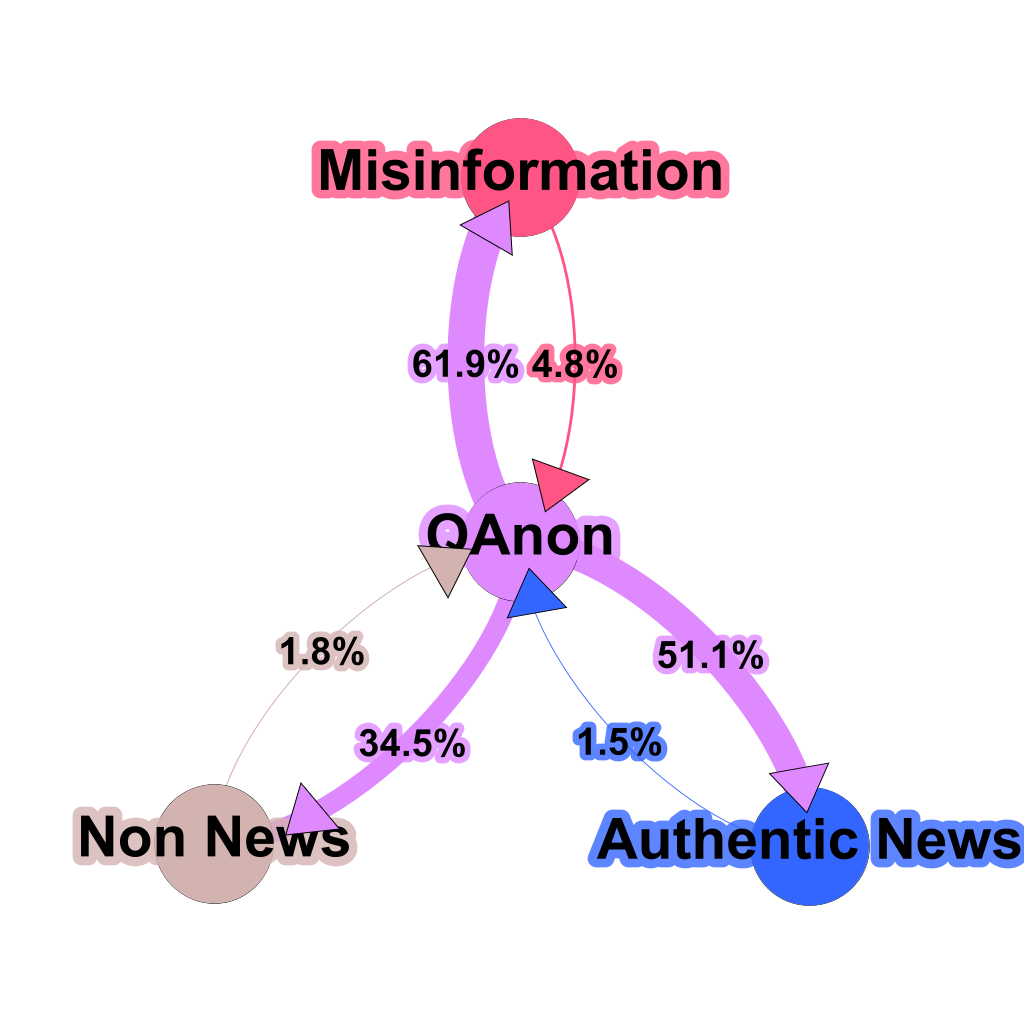}
  \caption{QAnon Connections}
\label{fig:qanon-sub3}
\hspace{2pt}
\end{subfigure}%
\begin{subfigure}{.30\textwidth}
  \centering
  \includegraphics[width=1\linewidth]{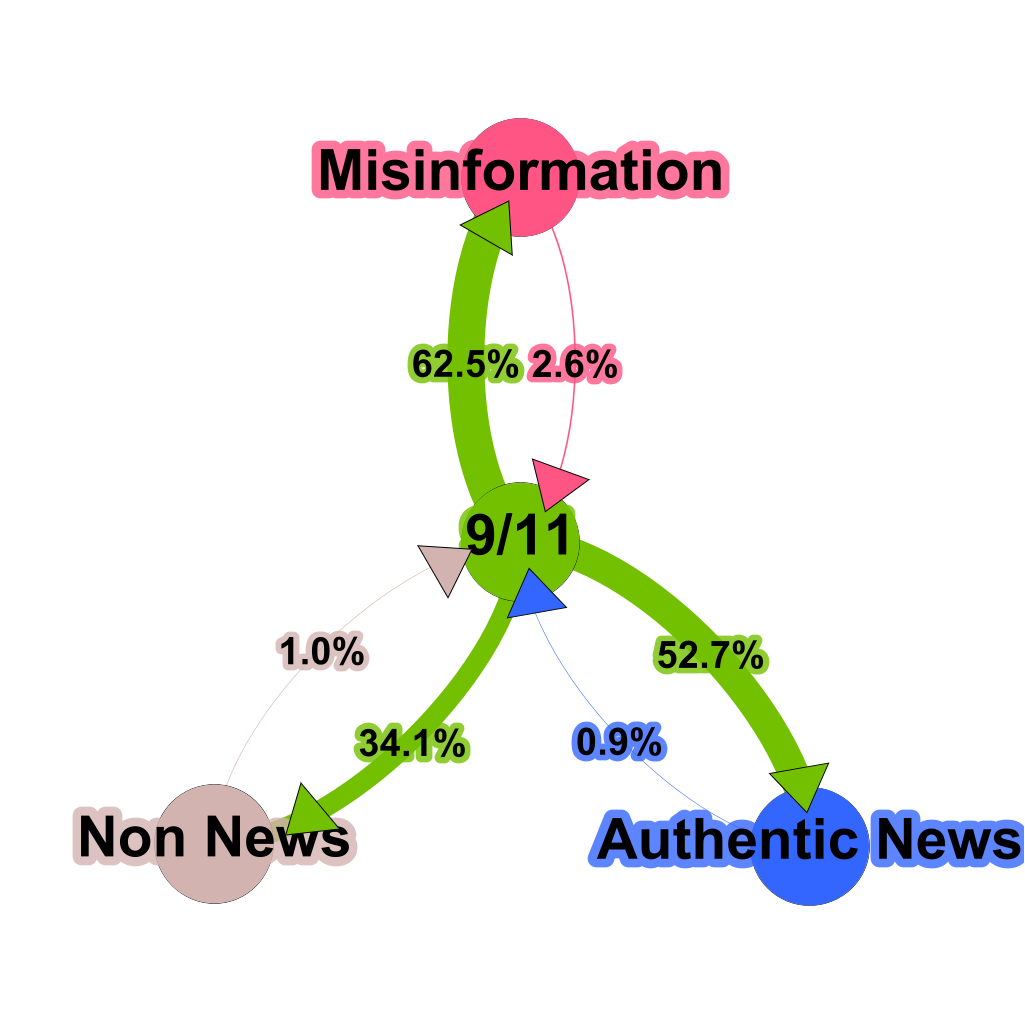}
    \caption{9/11 Connections}
  \label{fig:nine11-sub3}

\end{subfigure}

\begin{subfigure}{.30\textwidth}
  \centering
  \includegraphics[width=1\linewidth]{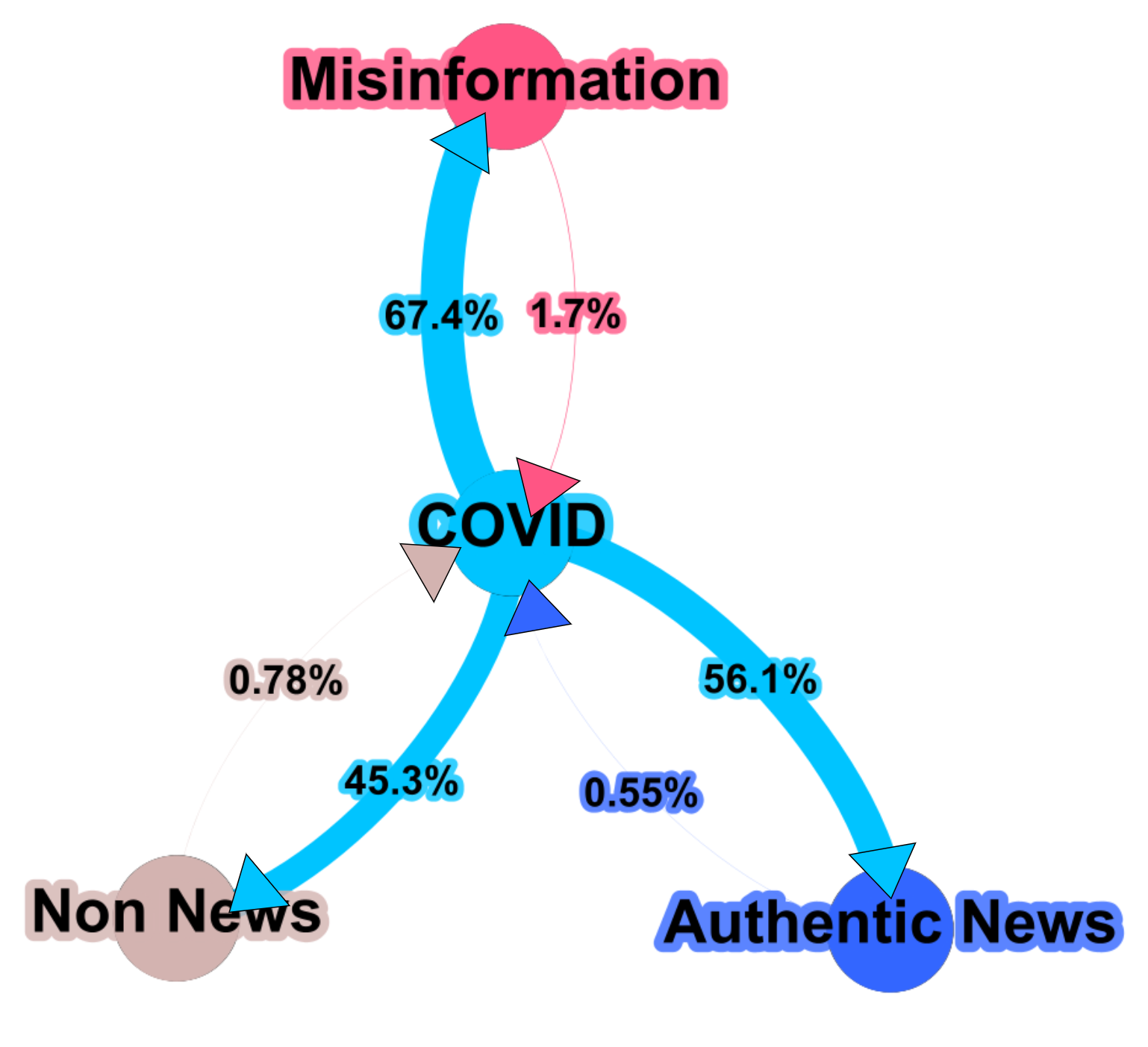}
   \caption{COVID Connections}
   \label{fig:covid-sub3}
\end{subfigure}
\hspace{2pt}
\begin{subfigure}{.30\textwidth}
  \centering
  \includegraphics[width=1\linewidth]{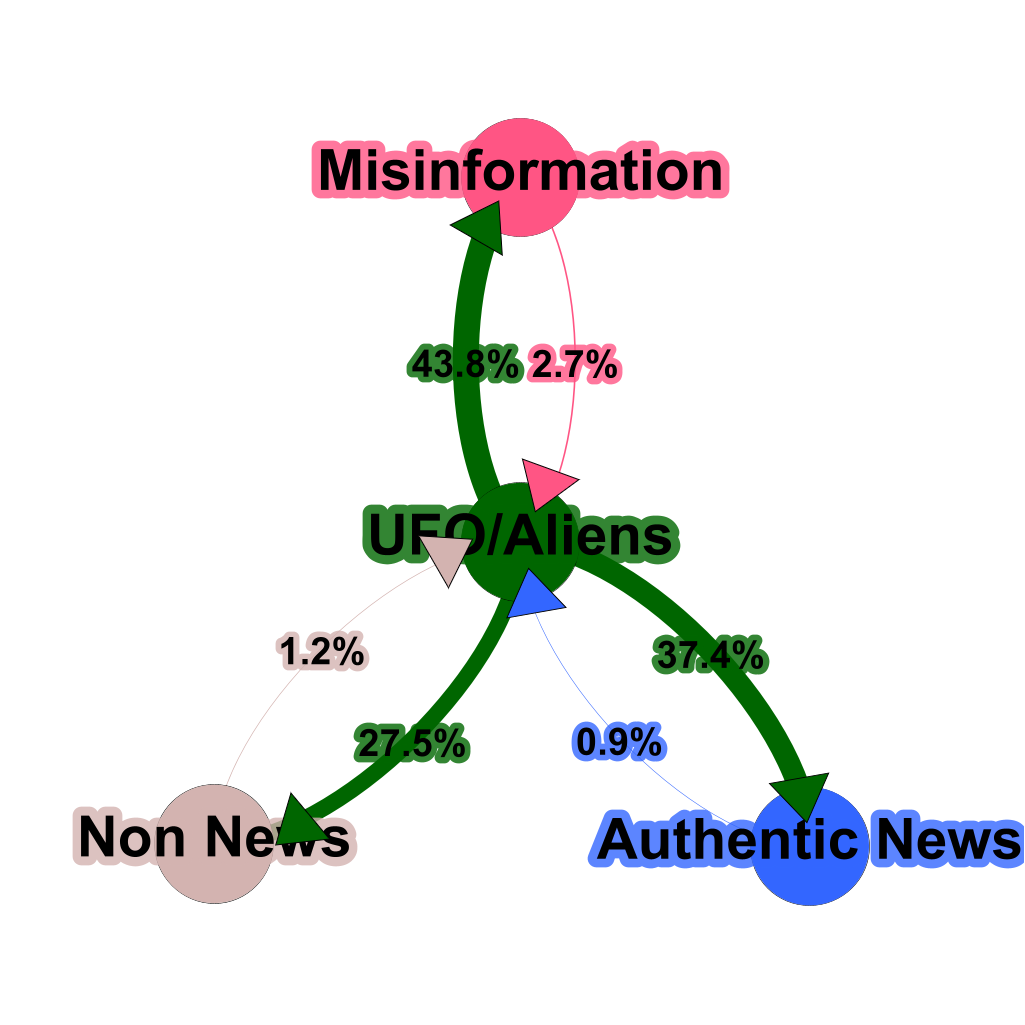}
   \caption{UFO Connections}
  \label{fig:ufo-sub3}
\end{subfigure}
\hspace{2pt}
\begin{subfigure}{.30\textwidth}
  \centering
  \includegraphics[width=1\linewidth]{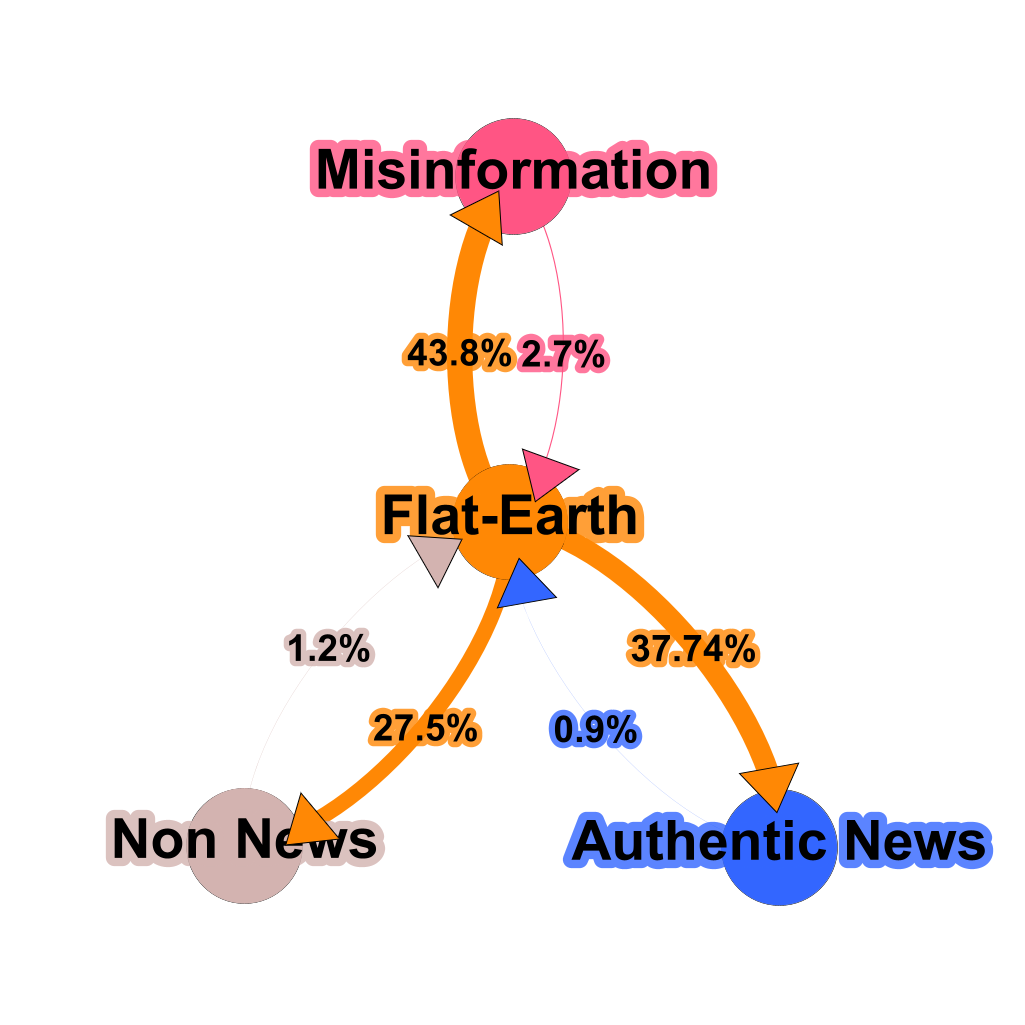}
   \caption{Flat-Earth Connections}
  \label{fig:flat-sub3}
  
\end{subfigure}
\caption{\textbf{Percentage of website domains hyperlinked by each website category that are also hyperlinked by another category}-- Our set of misinformation websites have the highest amount of shared domain connections with conspiracy theory websites. For instance, 4.8\% of all domains hyperlinked by misinformation websites are also hyperlinked by QAnon websites. By comparison, only at most 1.8\% and 1.5\% of all domains hyperlinked by non-news websites and authentic news websites, respectively, are also hyperlinked by QAnon websites.}
  \label{figure:conspiracies-to-different-types}
  \vspace{-15pt}
\end{figure*}
\subsection{Interactions between Conspiracies, the News Media, and  the Broader Web\label{sec:larger-conspiracy-interaction}}
Having given an overview of the conspiracy theory ecosystem, we now turn to understand these conspiracy theories' interactions with different parts of the Internet, and in particular, the news media. To understand the degree of interconnection between our conspiracy theory websites and different parts of the Internet, we again display in Figure~\ref{figure:conspiracies-to-different-types} the percentage of shared domain connections between our conspiracy theory categories and our set of authentic news, misinformation, and non-news-related websites. We now first investigate the percentage of news websites' hyperlinked domains that are shared by conspiracy theory websites before analyzing the percentage of conspiracy theory hyperlinked domains that are shared by news websites.


\textbf{\textit{Misinformation, Authentic News, and Non-News' Relationship with Different Conspiracy Theories:}}
As seen in Figure~\ref{figure:conspiracies-to-different-types}, non-news websites have relatively few connections with the same websites as conspiracy theories. All of the considered conspiracy theories have the least amount of shared connections with non-news websites. At the high end, only 1.8\% of the websites collectively hyperlinked by non-news websites are also hyperlinked by QAnon websites. We see similar behavior with authentic news websites. For every conspiracy theory, at the upper limit, only 1.5\% (QAnon) of authentic news collective connections are also hyperlinked by a conspiracy theory website category. This illustrates the degree to which both non-news and authentic news websites have maintained a distance between themselves and conspiracy theories. In contrast, we see our misinformation domains collectively have nearly triple the number of shared connections between themselves and every conspiracy theory website category. For example, while only 1.5\% of outward domain connections from authentic news websites are also hyperlinked by QAnon websites, 4.8\% of these connections are also hyperlinked by misinformation websites (Figure~\ref{fig:qanon-sub3}).
\begin{table}

    \small
    \centering
    \begin{tabularx}{.49\columnwidth}{Xrl}
    & Misinformation Outlets & \\
    \toprule
    Domain &  Unique URLs & Category  \\
    \midrule
    qmap.pub      &   44,502  & QAnon\\
    qntmpkts.keybase.pub & 22,567    & QAnon \\
    vaccineimpact.com   &   16,744& COVID\\
    thecovidblog.com&   1,668& COVID \\
    ae911truth.org  &   22,208 & 9/11 \\
    911research.wtc7.net & 19,970& 9/11 \\
    ufocasebook.com  & 7,293& UFO/Aliens\\
    ufodigest.com        &   2,706 & UFO/Aliens\\
     theflatearthsociety.org
     &   324& Flat-Earth
    \\
   flatrock.org.nz
 & 122& Flat-Earth \\
    \bottomrule
    \end{tabularx}
    \begin{tabularx}{0.49\columnwidth}{Xrl}
     & Authentic News Outlets & \\
    \toprule
    Domain &  Unique URLs & Category \\
    \midrule
    qposts.online
            &   63& QAnon \\
   prayingmedic.com
             &   29 & QAnon\\
      drrimatruthreports.com
 & 19& COVID \\
    drmalcolmkendrick.org
                &   6& COVID \\
    ae911truth.org
  & 1,228 & 9/11\\
      911truth.org
 & 1,037 & 9/11\\
   mufon.com        &   765 & UFO/Aliens\\
    ufosightingsdaily.com & 754 & UFO/Aliens \\

    theflatearthsociety.org
 & 126 & Flat-Earth \\
    wiki.tfes.org
 & 63 & Flat-Earth \\
    \bottomrule
    \end{tabularx}
   \caption{\textbf{Top Conspiracy Theory Websites Linked by Misinformation websites (left) and Authentic News websites (right) from Jan 2008 to Aug 2021}--- Misinformation websites link to a large amount of conspiracy theory outlets. We see very few connections to Flat-Earth websites. In contrast, authentic news websites rarely link to conspiracy theory websites across all categories of conspiracy theories.}
   \vspace{-10pt}
    \label{table:top_conspiracy_linked}
\end{table}

Examining the raw number of hyperlinks that misinformation and authentic news outlets have with specific conspiracy theory websites,  we again see that QAnon dominates, with the Flat-Earth conspiracy theory having among the fewest hyperlinks (Table~\ref{table:top_conspiracy_linked}) among misinformation domains. As seen in Table~\ref{table:top_conspiracy_linked}, 44K different hyperlinks from misinformation domains go to the ``QAnon Drop'' website qmap.pub. This website, in particular, acts as a reference for the QAnon conspiracy theory, aggregating and displaying all of QAnon's leader's messages in one place.  We note that while our set of authentic news domains, as seen in (Table~\ref{table:top_conspiracy_linked}) rarely hyperlink to conspiracy theory websites (especially for the COVID conspiracy theory) compared to misinformation domains, they do possess a somewhat stronger relationship with both 9/11 and UFO conspiracy theories. For example, for authentic news websites, the top linked domain is ae911truth.org, a website that provides ``evidence'' that the Twin Towers were part of a controlled demolition rather than a terrorist attack. This website, with its own Wikipedia page\footnote{
\url{https://en.wikipedia.org/wiki/Architects\_\%26_Engineers_for_9/11_Truth}}, Facebook Group\footnote{\url{https://www.facebook.com/ae911truthstudents/}}, and YouTube Channel\footnote{\url{https://www.youtube.com/user/ae911truth}}, while promoting a conspiracy theory, has apparently somewhat entered the mainstream.

\textit{\textbf{Conspiracy Theories Relationship with Misinformation, Authentic News, and Non-News Websites. The Extensive Use of News by Conspiracy Theories:}}
Conspiracy theories often attempt to interpret current events and use the resources utilized by misinformation and authentic news domains to do so (Figure~\ref{figure:conspiracies-to-different-types}). For both misinformation and authentic news, every conspiracy theory has heightened levels of shared domain connections compared with non-news websites. For every conspiracy category, at least 37.4\% of outward connections are also hyperlinked by authentic news websites, and at least 43.8\% of their outward connections are also hyperlinked by misinformation websites. This illustrates the high degree to which conspiracy theory websites, while ostensibly fringe, rely on websites and resources also utilized by more ``mainstream'' and popular websites.  Performing a series of Mann-Whitney U-tests comparing the relative levels of connections between every conspiracy theory website and mainstream and authentic news websites versus non-news websites, we find (after utilizing the Bonferonni correction for multiple statistical~\cite{armstrong2014use}) that every conspiracy theory group does indeed have elevated levels of shared domain connections with authentic news and misinformation websites.

\begin{table}
    \small
    \centering
    \begin{tabularx}{.32\columnwidth}{Xr}
    \toprule
   QAnon &  Unique URLs  \\
    \midrule
    zerohedge.com & 22,630 \\
    thegatewaypundit.com & 8,751 \\
    wikileaks.org & 5,727 \\
    theepochtimes.com & 5,502 \\
    breitbart.com & 5,454 \\
    rt.com & 5,452 \\
    veteranstoday.com & 3,511 \\
    bitchute.com & 2,552 \\
    freebeacon.com & 2,403\\
    sputniknews.com & 2,211 \\
    \bottomrule
    \end{tabularx}
    \begin{tabularx}{.32\columnwidth}{Xr}
    \toprule
    COVID  &  Unique URLs  \\
    \midrule
    healthimpactnews.com      &   2804\\
    bitchute.com     &   1,375 \\
    rt.com   &   1,135\\
    \mbox{off-guardian.org} & 707 \\
    corbettreport.com        &   698 \\
    \mbox{naturalnews.com} &   601 \\
    zerohedge.com & 596 \\
    greenmedinfo.com & 565 \\
    nomorefakenews.com & 490 \\
    globalresearch.ca & 425 \\
    \bottomrule
    \end{tabularx}
    
    \begin{tabularx}{.32\columnwidth}{Xr}
    \toprule
    9/11 &  Unique URLs  \\
    \midrule
    globalresearch.ca      &   2,082 \\
    prisonplanet.com    &   1,734 \\
    infowars.com   &   1,245 \\
    rawstory.com   &   901\\
    
    antiwar.com  &   855 \\
    veteranstoday.com & 839 \\
     wikileaks.org         &   799 \\
    \scriptsize{informationclearinghouse.info} & 750 \\
   whatreallyhappened.com & 684 \\
    rt.com      &   663\\
    
    \bottomrule
    \end{tabularx}
    \begin{tabularx}{.32\columnwidth}{Xr}
    \toprule
    Flat-Earth  &  Unique URLs  \\
    \midrule
    geoengineeringwatch.org & 67 \\
    breitbart.com     &   62 \\
    bitchute.com   &   47\\
    rt.com   &   33 \\
    thegatewaypundit.com &   31 \\
    naturalnews.com & 24 \\
    zerohedge.com & 18 \\
    rawstory.com & 16 \\
    globalresearch.ca & 13 \\
    abovetopsecret.com &12 \\
    \bottomrule
    \end{tabularx}
    \begin{tabularx}{.32\columnwidth}{Xr}
    \toprule
    UFO/Aliens &  Unique URLs  \\
    \midrule
    coasttocoastam.com      &   1,402 \\
    \mbox{collective-evolution.com}  &   275 \\
    rt.com     &   268 \\
   abovetopsecret.com & 143 \\
       sputniknews.com   &   130\\
        humansarefree.com & 118 \\
    disclose.tv      &   102 \\
     naturalnews.com       &   88 \\
    bitchute.com &  81 \\
    theepochtimes.com & 56 \\
    \bottomrule
    \end{tabularx}
   \caption{\textbf{Top misinformation websites hyperlinked by each category of conspiracy theory website}---Conspiracy theory websites heavily use misinformation websites. Many of these misinformation websites have been thoroughly described in prior works as promoting toxic ecosystems~\cite{hanley2021no,starbird2018ecosystem}.}
   \vspace{-10pt}
   \label{table:misinfo_top_link_in}
\end{table}

Looking at the specific domain-to-domain connections between our website lists, we further note that every conspiracy theory website category in our dataset also thoroughly utilizes our set of misinformation and authentic news websites. Each category collectively hyperlinks to at least 171~authentic news sites and at least 136~misinformation sites. Authentic news websites are more popular than misinformation, which may help explain this phenomenon. Using the Amazon Alexa Top List from March 1, 2021, we see that 255 (45.1\%) of the websites in our authentic news list are in the top 100K~websites while only 93 (17.5\%) of the misinformation are in the top 100K~\cite{amazon-top-mil}. Indeed, {nytimes.com}, {theguardian.com}, {cnn.com}, and {washingtonpost.com} are some of the most commonly hyperlinked authentic news sites by conspiracy theory websites (Table~\ref{table:top_link_in}). Similarly, popular misinformation sites like {rt.com}, {zerohedge.com}, {bitchute.com}, and {breitbart.com} whose role in promoting extreme beliefs~\cite{hanley2022happenstance,starbird2018ecosystem} are frequently hyperlinked by conspiracy theory websites (Table~\ref{table:misinfo_top_link_in}). 

\begin{table}
    \small
    \centering
    \begin{tabularx}{.32\columnwidth}{Xr}
    \toprule
     QAnon &  Unique URLs  \\
    \midrule
    cnn.com      &   15,606 \\
    foxnews.com &  7,980\\
    usatoday.com & 6,008 \\
    nytimes.com & 3,941 \\
    thehill.com & 3,203 \\
    theguardian.com & 2,747 \\
    cbsnews.com & 2,712 \\
    washingtonpost.com & 2,598 \\
    politico.com & 2,225 \\
    washingtonexaminer.com & 1,820 \\
    \bottomrule
    \end{tabularx}
    \begin{tabularx}{.32\columnwidth}{Xr}
    \toprule
      COVID &  Unique URLs  \\
    \midrule
    nytimes.com      &   753 \\
    theguardian.com     &   660\\
    cnn.com & 397 \\
    nejm.org & 354 \\
    sciencemag.org & 309\\
    forbes.com & 290 \\
    cnbc.com & 283 \\
    washingtonpost.com   &  254\\
    sciencedaily.com & 235 \\
    npr.org &221\\

    \bottomrule
    \end{tabularx}
    
    \begin{tabularx}{.32\columnwidth}{Xr}
    \toprule
      9/11 &  Unique URLs  \\
    \midrule
    nytimes.com      &   3,172 \\
    cnn.com   &   1,444\\
    washingtonpost.com     &   1,299 \\
    theguardian.com & 703 \\
    huffingtonpost.com & 645 \\
    salon.com & 465 \\
    examiner.com & 415 \\
    nydailynews.com & 357 \\
    usatoday.com  & 354 \\
    foxnews.com & 292 \\
    \bottomrule
    \end{tabularx}
    \begin{tabularx}{.32\columnwidth}{Xr}
    \toprule
      Flat-Earth  &  Unique URLs  \\
    \midrule
    nytimes.com      &   511 \\
     cnn.com  & 271 \\
         theatlantic.com &254 \\

    washingtonpost.com     &   235 \\
        nationalgeographic.com & 219 \\

       wired.com & 211 \\

    npr.org & 206 \\
        forbes.com  & 150 \\
    theguardiean.com  &143 \\
    newscientist.com &136 \\
    \bottomrule
    \end{tabularx}
    \begin{tabularx}{.32\columnwidth}{Xr}
    \toprule
      UFO/Aliens &  Unique URLs  \\
    \midrule
    examiner.com   &   528\\
    foxnews.com  &  511 \\
    nytimes.com & 470 \\
    huffingtonpost.com & 414 \\
    heraldtribune.com & 403 \\
    livescience.com & 389 \\
    popularmechanics.com & 336 \\
    cnn.com & 328 \\
    theguardian.com & 317 \\
    washingtonpost.com  & 309 \\
    \bottomrule
    \end{tabularx}
   \caption{\textbf{Top authentic news websites hyperlinked by each category of conspiracy theory website}--- Conspiracy theory websites make high use of authentic news websites. All conspiracy groups have {nytimes.com} and {theguardian.com} within their top ten hyperlinked authentic news websites.}
   \vspace{-10pt}
    \label{table:top_link_in}
\end{table}

\textit{\textbf{Centrality of Different Conspiracy Theories }}  Having examined the relationship between conspiracy theories and the news media, we finally seek to fully understand our set of conspiracy theories' role within the wider Internet. Utilizing Common Crawl network data~\cite{Nagel2021} over the indexed Internet (87.7 million websites), we thus determine the \textit{network centrality} of our set of conspiracy-focused websites to understand if each conspiracy theory website category is ``core'' (regularly utilized on the Internet) or ``peripheral''~\cite{hanley2021no,morina2022web,park2003hyperlink,easley2010networks,morina2022web}.

While only 446 of our conspiracy theory websites are within the Common Crawl dataset, this analysis allows us to fully understand the relative roles that each conspiracy theory website group in our dataset plays on the wider Internet. We further plot the centralities of our set of misinformation, authentic news, and non-news websites as references to compare our set of conspiracy theory centralities against. We note that while a widely used and often relied upon centrality measure, PageRank has been found to be susceptible to manipulation and spam~\cite{morina2022web}. As a result, to capture network centrality, we report a common revision of PageRank, namely the harmonic centrality, a measure that is also utilized by Common Crawl~\cite{rochat2009closeness,Nagel2021} and in other works~\cite{morina2022web}. As in Morina \textit{et~al.}~\cite{morina2022web}, for interpretability, we present the percentile of these measures and present the raw values as standardized z-scores (\textit{i.e.}, a value of 0 represents the average centrality and a value of 1 represents one standard deviation above the mean, \textit{etc.}).  We present our analysis utilizing PageRank and harmonic centralities for completeness.

\begin{figure*}
\begin{subfigure}{.48\textwidth}
  \centering
  \includegraphics[width=1\linewidth]{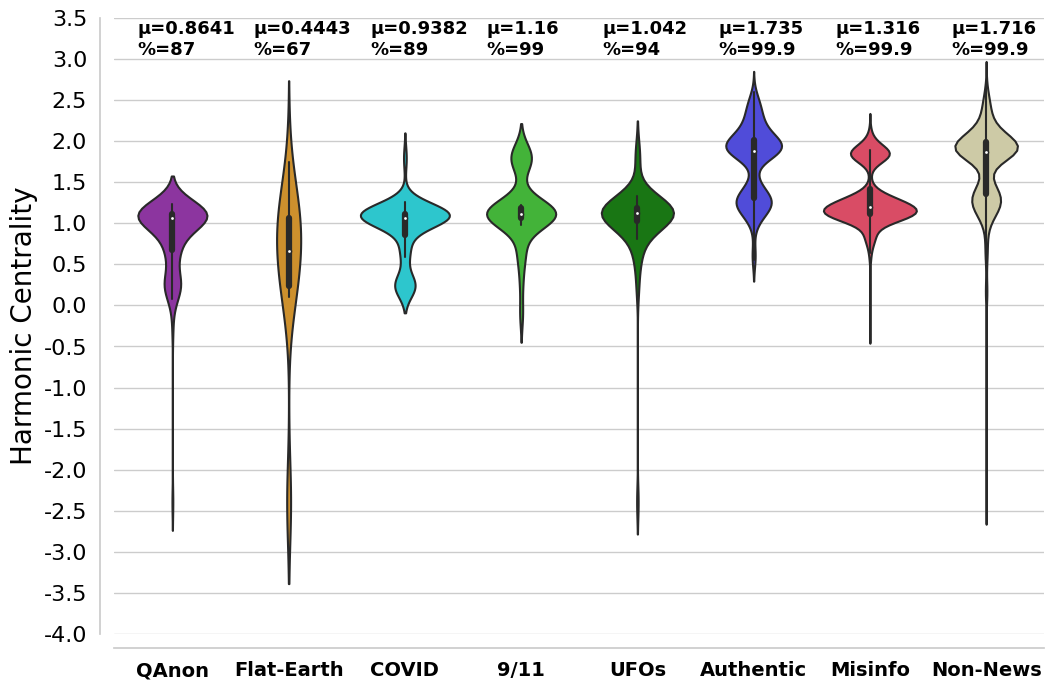}
\label{fig:harmonic-sub1}
\end{subfigure}
\begin{subfigure}{.50\textwidth}
  \centering
  \includegraphics[width=1\linewidth]{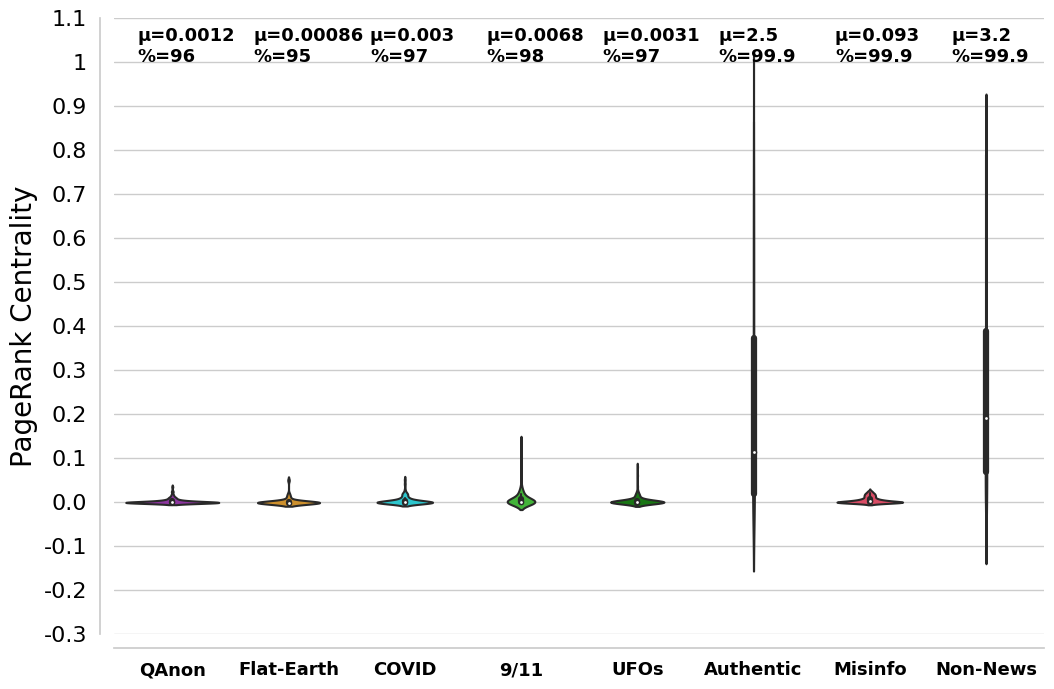}
  \label{fig:pagerank-sub2}
\end{subfigure}

\caption{\textbf{Average standardized harmonic and PageRank centrality measures with percentiles}--- Utilizing the full web crawl snapshot of Common Crawl from February 2021 to May 2021~\cite{Nagel2021}, we determine each conspiracy theory website group's harmonic and PageRank centralities. For reference, the small magazine {southfloridaopulence.com} has a harmonic centrality percentile of 78.48 while foxnews.com has a harmonic centrality percentile of 99.9997. Similarly, {southfloridaopulence.com} has a PageRank centrality of 92.700 while foxnews.com has a PageRank centrality percentile of 99.999.  For the mainstream, misinformation, and non-news PageRank centralities, we truncate the list to allow the distribution to fit on the same graph. As seen, each conspiracy theory group is on average more ``central'' to the web than the average website online by both methods of determining centrality.  }
\label{fig:conspiracy-domain-centrality-distributions}
\end{figure*}

As seen in Figure~\ref{fig:conspiracy-domain-centrality-distributions}, (from the set of conspiracy theory websites within Common Crawl data) each conspiracy theory website group had an above-average centrality on the Internet between February and May of 2021. Namely, each website group was more central than the average and median website on the Internet. Furthermore, each group of conspiracy theory websites has domains that are \emph{more} central than many authentic news, misinformation, and nonnews websites. We note that this is despite many of our conspiracy theory websites not appearing in the Amazon Alexa top million continuously during this period (Table~\ref{table:conspiracy-hubs})~\cite{amazon-top-mil}. While as a whole this could be due to our biased selection of websites (Section~\ref{sec:conspiracy-focused-dataset}), this result \textit{does} illustrate the degree to which several conspiracy theories websites and their associated theories have been utilized and become central to the Internet. Specifically, as seen in Table~\ref{table:conspiracy-hubs}, each conspiracy theory had multiple websites within the top 99\% percentile of centralities on the Internet.

\begin{table}
\centering
\small
\begin{tabular}{lllll}
\toprule
{Category} &{Domain} & {Harmonic w/ Perc.} & {PageRank w/ Perc. } & Alexa Rank   \\
\midrule
QAnon &qactus.fr           &   1.230 (99.49\%) 
    &   0.0106 (98.49\%) & 265,967 \\
QAnon &qmap.pub             &   1.210 (99.40\%) 
    &   0.0174 (99.32\%) & ---
 \\

 Flat-Earth &theflatearthpodcast.com               &1.743 (99.86\%) 
   &  0.0139  (99.16\%) & ---
    \\
    
     Flat-Earth &nasalies.org &   1.712 (99.79\%) 
    &    0.0101 (99.82\%)  & ----
    \\
COVID &vaccineimpact.com                    & 1.818 (99.96\%) 
   &   0.0508  (99.77\%) & ---
    \\
COVID &plandemicseries.com          &   1.768  (99.90\%)  
  &  0.0163 (99.28\%)  & 244,082
     \\
UFO/Aliens  &mufon.com               &  1.858  (99.98\%)    &   0.0798  (99.85\%)   & ---
     \\
    UFO/Aliens & ufosightingsdaily.com               &  1.830  (99.97\%)   &
 0.0603   (99.80\%)  & --- \\
9/11  & ae911truth.org   &    1.872 (99.98\%)  & 0.1330 (99.91\%)  & ---
    \\
9/11  &patriotsquestion911.com  &   1.820 (99.96\%)   &   0.0267 (99.56\%)   & ---
     \\
\bottomrule
\end{tabular}
\caption{\textbf{Most central domains per conspiracy-focused category }---The top two most central two websites across the Common Crawled Internet from February 2021 to May 2021~\cite{Nagel2021}, in each conspiracy theory category. The harmonic and PageRank centralities of each website are reported as z-scores with their respective percentile. The Amazon Alexa rank is given as of March 1, 2021~\cite{amazon-top-mil}. 
}
\label{table:conspiracy-hubs}
\vspace{-10pt}
\end{table}

\subsection{Summary}
In this section, we documented the complex role conspiracy theories have had with the wider Internet and, in particular, the news media. First, we showed the interdependence among our conspiracy theories and that QAnon remains a prominent theory, maintaining a higher-than-average connection with our other conspiracy theories. 
 While each group of conspiracy theory websites in our dataset was distinct, we find that every group hyperlinks to at least 7.5\% of the same domains as every other group, with QAnon even connecting to 35.6\% of the same domains as COVID websites (Table~\ref{table:conspiracy-domain-connections}). This is in contrast to the misinformation, authentic news,  and non-news domain which share at most 4.8\% (QAnon) with any of our conspiracy theory website groups and as little as 0.55\% (COVID) (Figure~\ref{figure:conspiracies-to-different-types}). Our results largely accord with news reporting that QAnon has become a ``big tent conspiracy theory'' that constantly incorporates new claims and theories~\cite{Roose2021}. We further find that all of our conspiracy theories extensively hyperlink and utilize news media, and in particular misinformation outlets within their online content. We found that every conspiracy theory category in our dataset has significant percentages of shared domain connections with both authentic news and misinformation domains. Thus while promoting fringe ideas, we see that each of the conspiracy theory categories websites still rely on and hyperlink to many of the same websites that more ``mainstream'' websites also utilize. However, as expected, we find that out of the three categories of misinformation, authentic news, and non-news, misinformation websites have by far the most connections with our set of conspiracy theory websites. These misinformation websites have nearly 3 times the percentage of shared domain connections with conspiracy theory websites compared to authentic news and non-news websites.

We further documented that while our sets of conspiracy theory websites are considered fringe, each of them has websites that occupy a ``core'' place on the Internet. Each conspiracy theory category that we document has websites that are within the top 99\% percentile of centrality on the Internet according to both harmonic and PageRank centrality measures.

\section{RQ2: Conspiracy Theories and the News Media over Time}
\label{sec:temporal}

Having outlined the static interactions between conspiracy theories and news media,  we now discuss the 
 \textbf{dynamic} interactions between online conspiracy theories and the news media ecosystem (and in particular misinformation). Previous social science research has indicated that conspiratorial thinking has remained constant throughout much of human history~\cite{uscinski2014american,van2014power}, spiking temporarily only during moments of social unrest. Given that (1) political polarization and social upheaval have increased significantly since 1994 in the United States~\cite{PewPolarization} and (2) having determined that misinformation and authentic news both play a significant role in the conspiracy theory ecosystem, we now measure (1) whether the popularity of conspiracy theories have increased during the past decade and (2) if news sites have had an increasing role within on the conspiracy-theory ecosystem. 

\paragraph{Conspiracy-Oriented Websites}
For the rest of this paper, we define \textit{conspiracy-oriented} websites as websites that have more connections from conspiracy theory websites than from authentic news and non-news websites (\textit{i.e.}, the majority of a site's inward links in a domain-based graph are from conspiracy websites [excluding misinformation domains]).  We use this definition to understand how misinformation and authentic news interact with \textbf{conspiracy-related materials generally} rather than just our 755~websites. Given that conspiracy theory websites are smaller and have both fewer pages and hyperlinks to other websites, this is a conservative definition. Indeed, the median conspiracy theory site links to 73~other domains, the median non-news site to 176~sites, and the median authentic news website to 2,202~sites. In addition, while there are 755~conspiracy theory websites in our dataset, there are a total of 1107~authentic news and non-news websites. Out of the 3.8 million different domains that were hyperlinked in our scrapes, 116K (3.05\%) are \textit{conspiracy-oriented}. 207~of our misinformation domains overlap with our \textit{conspiracy-oriented} domains, including websites like {zerohedge.com}, {gulagbound.com}, and {prisonplanet.tv}. All of our conspiracy theory websites are considered \textit{conspiracy-oriented}. We finally note that this definition removed fairly common popular websites like {twitter.com}, {facebook.com}, {reddit.com}, and {nytimes.com}.

\subsection{The Web's Changing Relationship with Different Conspiracy Theories}\label{sec:over-time}
\begin{figure}
\centering
\includegraphics[width=\columnwidth]{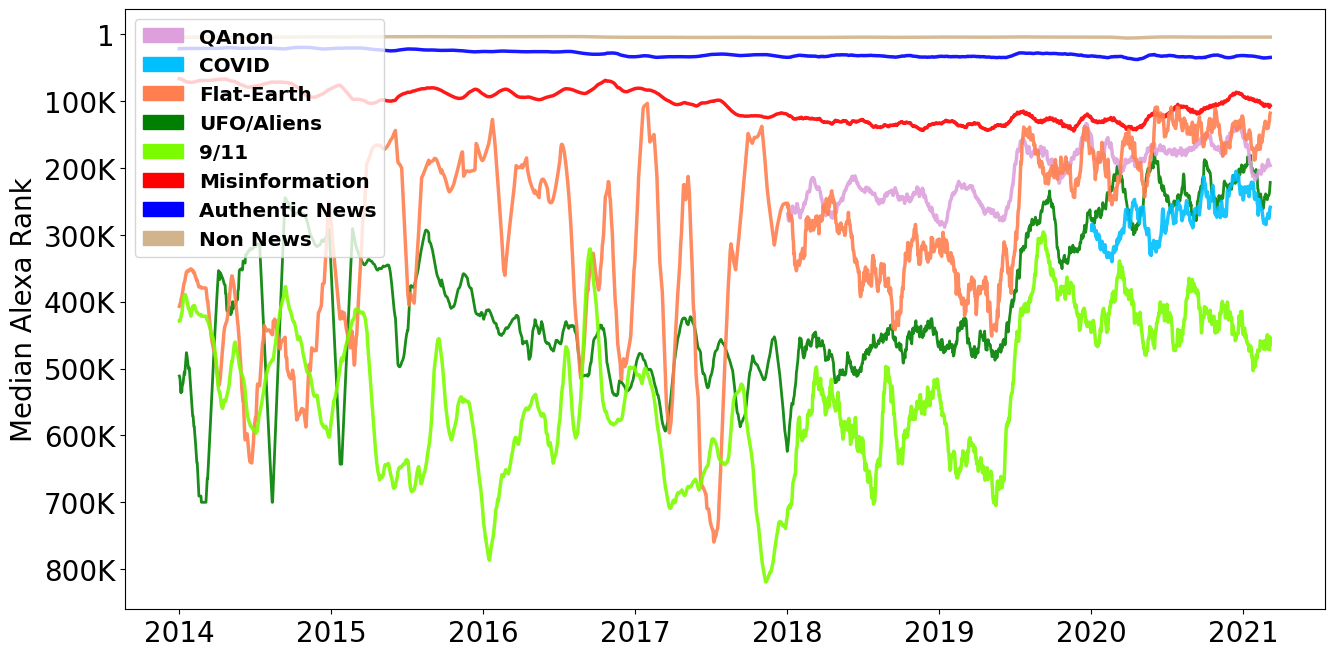} 
\caption{\textbf{30-Day moving average of the median Amazon Alexa rank of non-news, authentic news, misinformation, and each category of conspiracy theory website}--- We note a large increase in the median popularity of all conspiracy theories website categories around August 2019 (we ensured that this trend is specific to these groups of websites by plotting the median Amazon Alexa rank of 500 random websites, validating that the rank of the set of websites was stable and did not exhibit the increase seen for the conspiracy theory websites). Note: Amazon Alexa changed its methodology for calculating rankings on January 30, 2018~\cite{scheitle2018long, Tajalizadehkhoob2019}.}
\label{figure:popularity_over_time}
\end{figure}


To begin, we first investigate the relative website popularity of authentic news, misinformation, non-news, and each conspiracy theory group over time. To do this, we consider each group's median rank in the Alexa Top Million list~\cite{amazon-top-mil}. We note that Amazon Alexa changed its methodology for calculating rankings on January 30, 2018~\cite{scheitle2018long, Tajalizadehkhoob2019}. Previously, Alexa rankings were averaged over three months of data from URLs visited by users with a given browser extension installed. While the exact change was not announced by Amazon, it was found that the rankings after January 30, 2018, were not averaged over multiple days~\cite{Tajalizadehkhoob2019}.

Between 2014--2021, the median Alexa ranks of authentic news, misinformation, and non-news websites have all remained relatively stable, while conspiracy theory websites' popularity has varied widely (Figure~\ref{figure:popularity_over_time}). Separately, however, all conspiracy theories experienced a dramatic increase in popularity between July and August 2019. While we cannot show definitive causality, we note that this coincides with the El Paso, Texas, and Dayton, Ohio, mass shootings that took place on August 3rd and 4th, respectively. In both cases, the shooters posted manifestos on the QAnon-associated website {8chan.net} (not in our dataset, as it is the previous form of 8kun). As previously noted, QAnon material and QAnon websites like 8chan often host and share material for other conspiracy theories; the large simultaneous increase in popularity among each conspiracy website group largely follows from QAnon acting as a means by which people access other conspiracy theories~\cite{Roose2021}. We finally note that following these events, {8chan.net} shut down, subsequently rebranding as {8kun.top}. Since this increase in popularity, the median popularity of each conspiracy theory website group has remained high, with the exception of the 9/11 conspiracy theory (Figure~\ref{figure:popularity_over_time}). We ensured that these trends were specific to these groups of websites by plotting the median Alexa rank of 500 random websites from the Alexa list and validating that the rank of the set of websites was stable and did not exhibit the increase seen for the conspiracy theory websites.  

\begin{figure}
\begin{subfigure}{1.0\textwidth}
\centering
\includegraphics[width=\columnwidth]{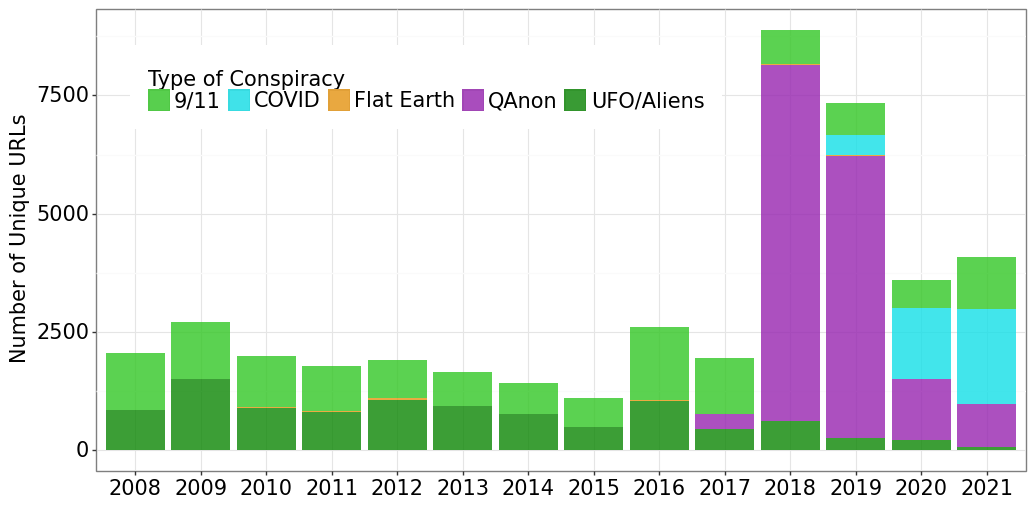} 
\end{subfigure}
\caption{\textbf{Hyperlinks from misinformation websites to individual conspiracy theory website groups}--- (Note: Hyperlinks were only collected to Aug 2021) The vast majority of links from news websites to conspiracy theory websites are from misinformation sites.  After the appearance of QAnon, there was a massive increase in links to conspiracy theory websites. The decline of QAnon was followed by a large increase in COVID conspiracy theory website links.}
\label{figure:misinfo-main-conspiracy}
\vspace{-10pt}
\end{figure}

\subsection{The Misinformation and Authentic News Changing Relationship with Different Conspiracy Theories}\label{sec:misinfo-authentic-time}
\begin{figure}
\begin{subfigure}{1.0\textwidth}
\centering
\includegraphics[width=\columnwidth]{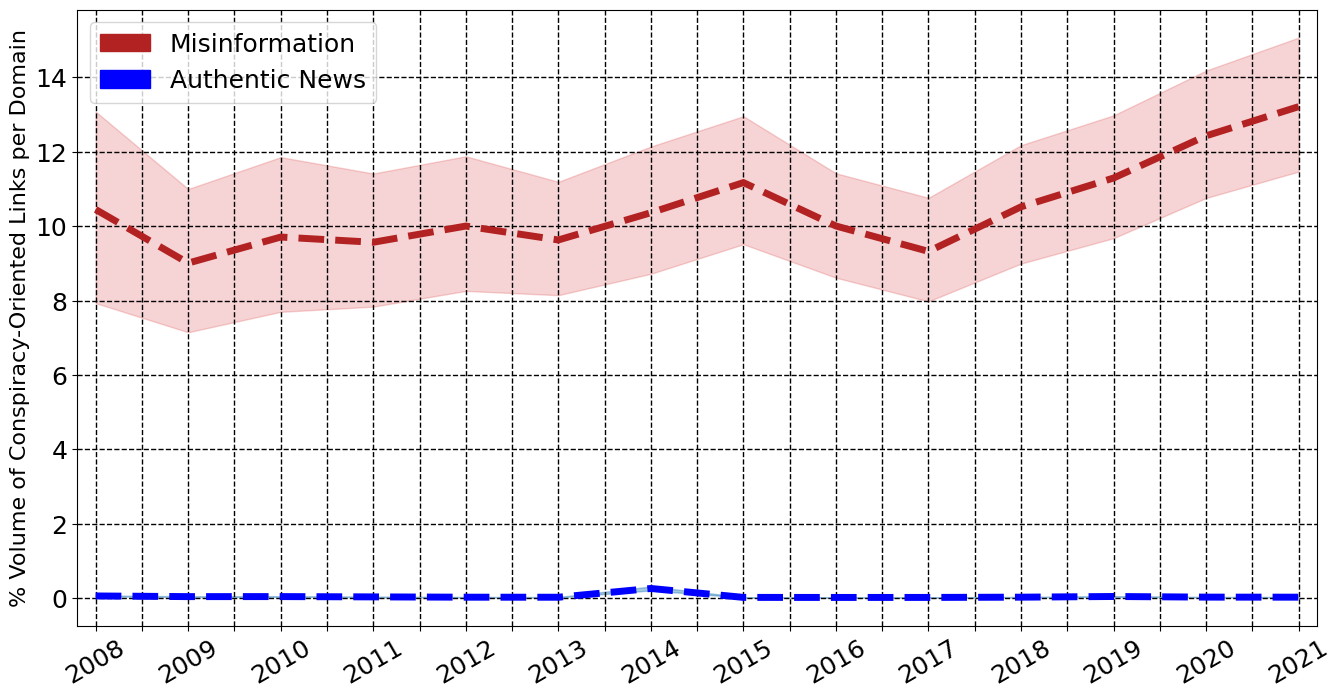} 
\end{subfigure}
\caption{\textbf{Average percentage of hyperlinks to \textit{conspiracy-oriented} websites from misinformation and authentic news websites}--- Plotted is the average percentage per source domain with 95\% bootstrap confidence intervals of hyperlinks to conspiracy-oriented websites from misinformation and authentic news websites. Misinformation websites have increasingly linked their users to \textit{conspiracy-oriented} domains since 2008. Authentic news websites have refrained from directing their users to websites that promote conspiracy theories.}
\label{figure:time-series-change}
\end{figure}

Next looking at the interaction over time between news websites and conspiracy theory websites we see that misinformation websites have increasingly hyperlinked to conspiracy theory and \textit{conspiracy-oriented} websites, concurrent with the increase in popularity of conspiracy theory websites (Figures~\ref{figure:popularity_over_time},~\ref{figure:misinfo-main-conspiracy}, and~\ref{figure:time-series-change}). We specifically observe a massive jump in the number of hyperlinks to conspiracy theories due to the arrival of QAnon in 2017--2018 (Figure~\ref{figure:misinfo-main-conspiracy}). This was later supplemented by COVID hyperlinks at the start of the COVID pandemic in late 2019 and early 2020. We see relatively few Flat-Earth links; this may suggest its somewhat separate role within the conspiracy theory ecosystem. This lack of links further matches the lower centrality of Flat-Earth websites that we observed in Section~\ref{sec:larger-conspiracy-interaction} in Table~\ref{table:conspiracy-hubs} and Figure~\ref{fig:conspiracy-domain-centrality-distributions}.

Between 2008 and 2021, looking at the average percentage of \textit{conspiracy-oriented} external hyperlinks per source domain (especially compared to authentic new websites) misinformation websites have hyperlinked to more and more \textit{conspiracy-oriented} domains. This percentage captures the growth of news media to conspiracy-oriented materials online have increased in general rather than to just our five distinct conspiracy theories. This peaked in 2021 at 13.2\% from a low of 9.01\% in 2009, a 46.6\% relative increase. Looking at all external hyperlinks from misinformation websites, instead of a per domain average, we see that the percentage that goes to \textit{conspiracy-oriented} starts from a low of 13.9\% in 2008 and increases to a high of 19.1\% in 2021, a 37.8\% relative increase (Figure~\ref{figure:time-series-change}). Looking at some of the top \textit{conspiracy-oriented domains}, blazetv.com is one of the most commonly hyperlinked. Blazetv.com is a television network founded by conservative commentator Glenn Beck~\cite{ahmad2022analysis}, which has long been known to spread misinformation. For example, the network spread a debunked claim that 1000 mail-in ballots were found in a dumpster in California in September 2020~\cite{benaissa2021sources}. Some of the other most commonly linked \textit{conspriacy-oriented} websites are Gab, Parler, and MeWe (Table~\ref{table:conor_top_link_in}). All three sites are known to spread disinformation, conspiracy theories, and hate~\cite{mathew2019spread,aliapoulios2021early,myers2021propaganda}. Parler was even utilized as a communication forum by rioters who attacked the United States Capitol on January 6, 2021~\cite{Benner2021}.

\begin{table}
    \small
    \centering
    \begin{tabularx}{.49\columnwidth}{Xr}
    \toprule
    Domain &  Unique Hyperlinks  \\
    \midrule
    blazetv.com   &   3,308,119 \\
    theabovenetwork.com    &   2,581,659\\
    mitocopper.com  &   2,093,941\\
    russian-faith.com  &   1,445,839 \\
    parler.com&   1,170.825 \\
    banned.video & 931,672\\
    zerohedge.com& 808,906 \\
    gab.com        &   671,900 \\
    gab.ai       &   659,738\\
    mewe.com & 622,483 \\
    \bottomrule
    \end{tabularx}
    \begin{tabularx}{0.49\columnwidth}{Xr}
    \toprule
    Domain &  Unique Hyperlinks  \\
    \midrule
    ok.ru             &   59,803 \\
    9nl.us & 8,432 \\
    rt.com             &   8,169 \\
    zerohedge.com       &   1,559 \\
    universalpressrelease.com & 1,312 \\
    thegatewaypundit.com & 933 \\
    paulcraigroberts.org & 920 \\
    christchurch.org.nz & 908 \\
    antiwar.com & 783 \\
    sputniknews.com & 768 \\
    \bottomrule
    \end{tabularx}
    \caption{\textbf{Top \textit{conspiracy-oriented} websites hyperlinked by misinformation websites (left) and authentic news websites (right) between January 2008 and August 2021}--- Misinformation websites hyperlink to a large number of websites normally considered to promote conspiracy theories and misinformation, most prominently Gab, Parler, and MeWe~\cite{mathew2019spread,aliapoulios2021early,myers2021propaganda}. Authentic news websites largely do not direct users to conspiracy-related domains but do occasionally link to Russian-operated websites that are known to spread disinformation.}
   \vspace{-10pt}
   \label{table:conor_top_link_in}
\end{table}

Even considering the larger set of 116K conspiracy-oriented domains, authentic news domains have largely avoided linking their audience to conspiratorial material, with the notable exception of some Russian websites. As seen in Table~\ref{table:conor_top_link_in}, the top \textit{conspiracy-oriented} domains that authentic news websites hyperlink are primarily Russian websites. {Ok.ru} is a Russian social media second in popularity in Russia, behind the Russian social media website VK/VKontakte, and just ahead of Facebook~\cite{OkSprinklr2017}. {Rt.com} is the website of the Russian Television (RT) network, known to spread disinformation~\cite{starbird2018ecosystem}. As documented by Yablokov \textit{et~al.}~\cite{yablokov2021russia}, RT and the Russian government have consistently engaged with conspiracy theories as a form of public diplomacy. 199 different authentic news websites hyperlink to articles from rt.com including nytimes.com and washingtonpost.com. However, even including this wider set of   \textit{conspiracy-oriented} websites, the percentage of \textit{conspiracy-oriented} hyperlinks posted by authentic news sites is negligible (Figure~\ref{figure:time-series-change}).

\section{RQ3: The Role of Misinformation and Authentic News in the Popularity of Conspiracy Theories}

In the last section, we showed that misinformation websites frequently hyperlink to conspiracy theories, whereas authentic news websites rarely do. We now analyze whether the behaviors of misinformation and authentic news websites have actively encouraged/discouraged the popularity of conspiracy theories. Specifically, we determine how changes in the volume of hyperlinks to conspiracy theory websites from misinformation websites have correlated with the popularity of our conspiracy theory domains.  Similarly, (given the relative lack of hyperlinks to our conspiracy theory websites from authentic news domains), we determine the relationship between the popularity of conspiracy theory websites and authentic news websites' frequency of mentioning different conspiracies.

\begin{figure}
\centering
\begin{minipage}[c]{0.5\textwidth}
\includegraphics[width=\columnwidth]{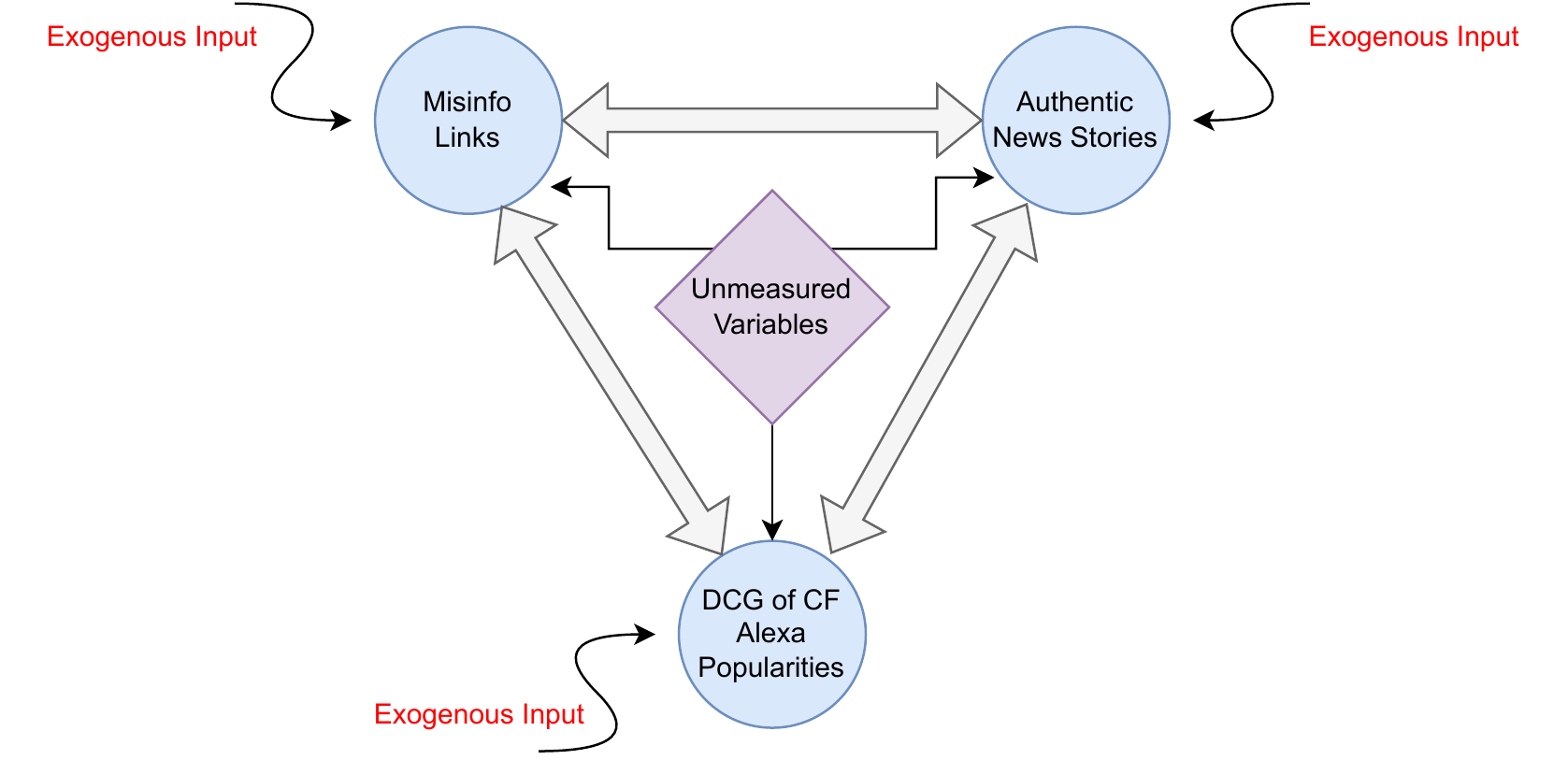}
\end{minipage}
\begin{minipage}[c]{0.45\textwidth}
\caption{\textbf{Model to determine partial Granger causality relationships between misinformation hyperlinks, authentic news stories, and the popularity of conspiracy-focused websites}--- We model the influence of misinformation website hyperlinks and authentic news stories about conspiracy theories on the popularity of our set of conspiracy focused websites~\cite{amazon-top-mil}.}
\label{figure:full-model}

\end{minipage}
\end{figure}
\begin{figure*}
\begin{subfigure}{.49\textwidth}
  \centering
  \includegraphics[width=1\linewidth]{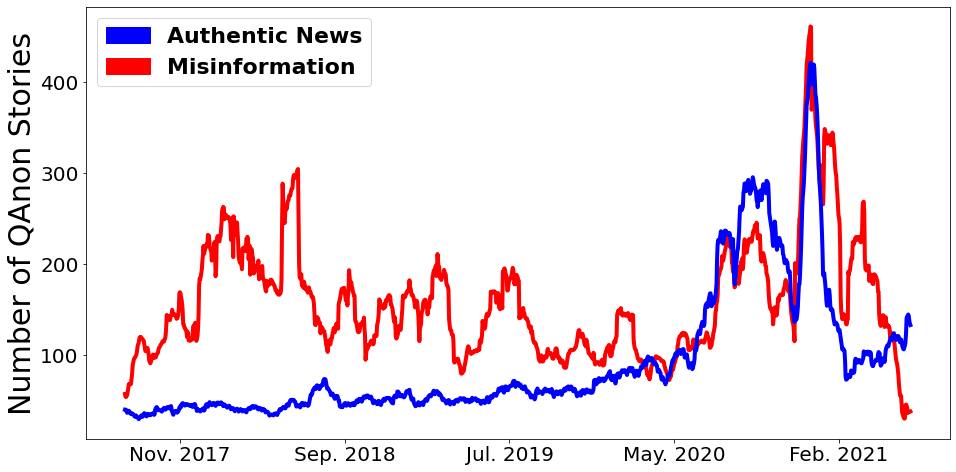}
  \caption{QAnon}
\label{fig:qanon-stories-sub1}
\end{subfigure}%
\begin{subfigure}{.49\textwidth}
  \centering
  \includegraphics[width=1\linewidth]{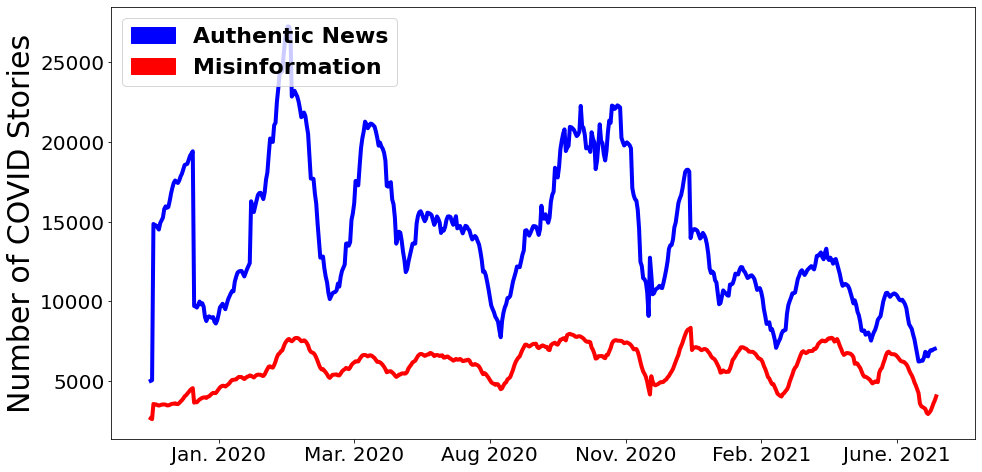}
  \caption{COVID}
  \label{fig:covid-stories-sub2}
\end{subfigure}
\vspace{8pt}
\begin{subfigure}{.49\textwidth}
  \centering
  \includegraphics[width=1\linewidth]{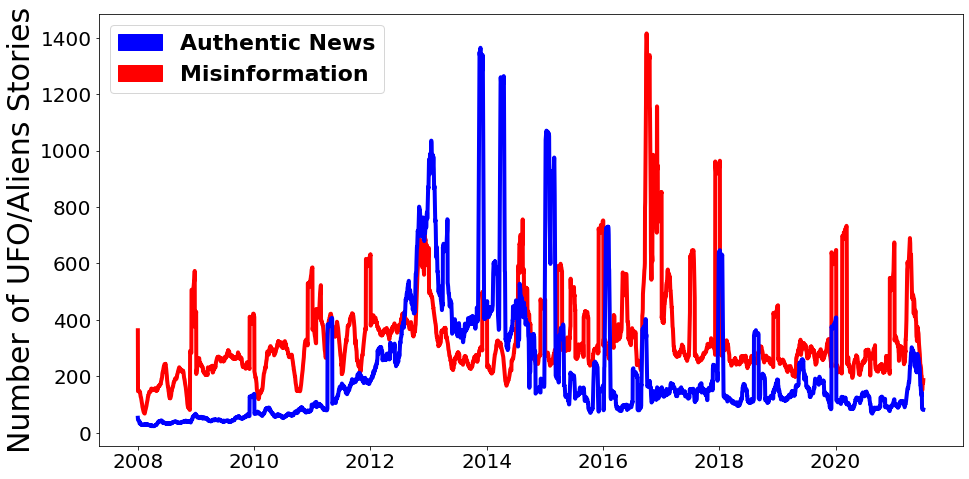}
  \caption{UFO/Aliens}
  \label{fig:ufo-stories-sub2}
\end{subfigure}
\begin{subfigure}{.49\textwidth}
  \centering
  \includegraphics[width=1\linewidth]{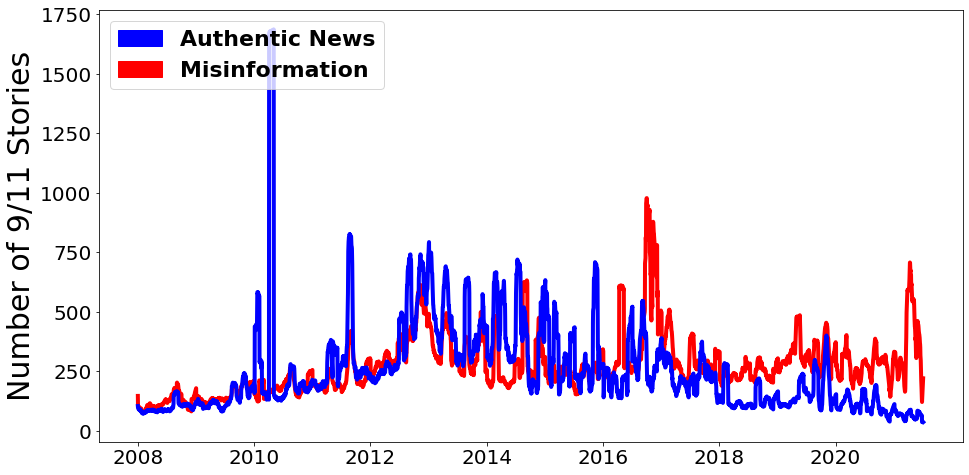}
  \caption{9/11}
  \label{fig:nine11-stories-sub2}
\end{subfigure}
\vspace{5pt}
\begin{subfigure}{.49\textwidth}
  \centering
  \includegraphics[width=1\linewidth]{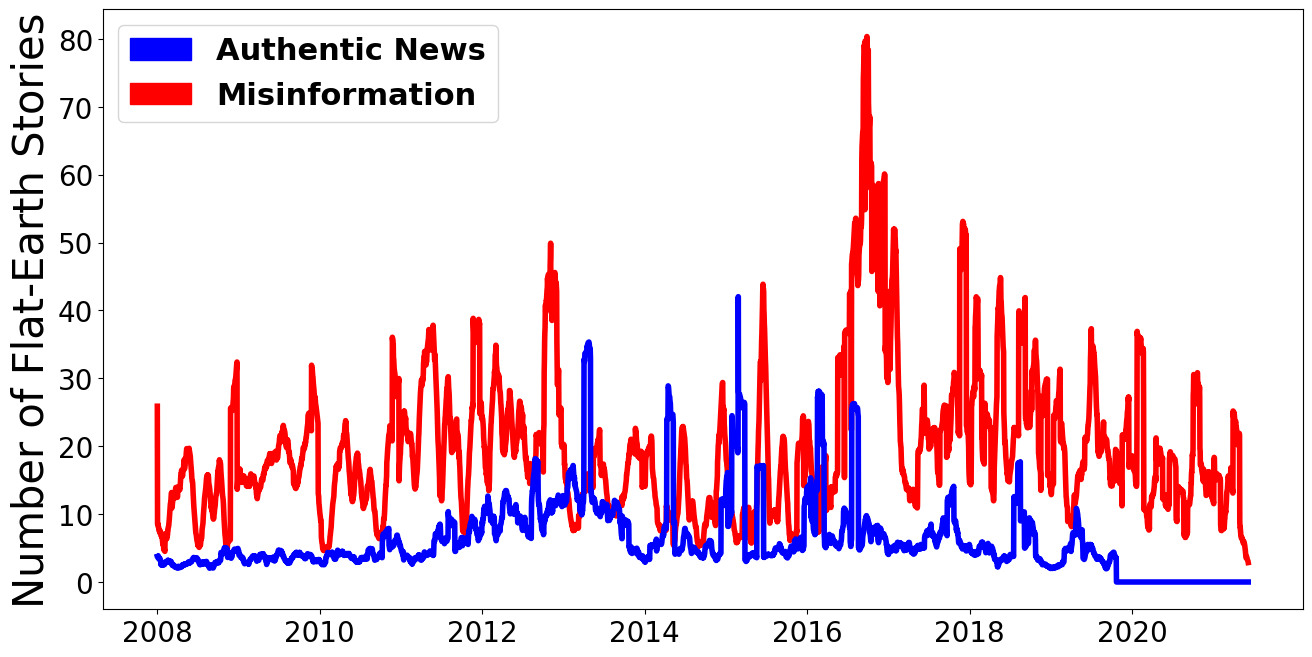}
  \caption{Flat Earth}
  \label{fig:flat-stories-sub2}
\end{subfigure}
\caption{\textbf{News stories mentioning each conspiracy theory}--- We show the number of misinformation and authentic news stories that mention each conspiracy theory in our study.
}
\label{fig:conspiracy-num-stories}
\vspace{-10pt}
\end{figure*}

\textit{\textbf{Setup.}} To examine these relationships, we utilize the notion of \textit{partial Granger causality} outlined in Section~\ref{sec:methodology-granger}. For full details on partial Granger causality see Appendix~\ref{sec:p-granger-causality}.\footnote{To perform tests for Granger-causality, analyzed time series must be \textit{stationary}. Stationarity implies that several statistics (\textit{i.e. mean}, variance) of the time series data must not change over time. Our untransformed data is not stationary given the large changes it makes over time (Figures~\ref{figure:popularity_over_time},~\ref{figure:misinfo-main-conspiracy}, and \ref{fig:conspiracy-num-stories}). We perform \textit{difference transformation}, performing cross-checks against the Augmented Dickey-Fuller (ADF) test and the Kwiatkowski–Phillips–Schmidt–Shin (KPSS) test to ensure stationarity. For all time-series data that we analyze, each series rejects the null hypothesis in the ADF test (implying stationarity) and accepts the null hypothesis in the KPSS test (also implying stationarity). To pick the appropriate lag (\textit{i.e.}, the number of past values to consider in the autoregressive model), we minimize the Bayes Information Criterion (BIC). We further note that to perform tests for Granger-causality, after fitting the autoregressive model, there must not be serial correlations among the residuals after fitting. Serial correlations imply that there are leftover patterns that are not explained by the model. For each model that we fit, we apply the Durbin-Watson test to test for serial correlations. For each model we fit, we find, utilizing this test, no evidence of serial correlations.} Specifically, utilizing this construct, we determine whether the number of hyperlinks to each specific category of conspiracy theory websites from misinformation websites has a Granger-causal influence on the popularity of this same category of conspiracy theory websites. This is also while taking into account the time-dependent influence of authentic news websites writing about the considered conspiracy theory as well as the influence of potential environmental exogenous inputs and endogenous latent variables. Conversely, we also determine whether the number of stories about a given conspiracy theory from authentic news websites has a partial Granger-causal influence on the popularity of this same conspiracy theory while also taking into account the time-dependent influence of hyperlinks from misinformation websites and potential environmental exogenous input and endogenous latent variables. For our experiment, we note that the noise variables modeled would represent potentially erroneously documented links and stories, as well as ranks that were incorrect within the Amazon Alexa top list. We picture this model in Figure~\ref{figure:full-model}.

As a proxy for the popularity of each conspiracy theory, we utilize a binary version of the discounted cumulative gain (DCG) of the daily Amazon Alexa ranks. DCG is a measure of the quality of rankings~\cite{jarvelin2002cumulated,chen2018investigating}. The binary version of DCG is calculated as follows:
\begin{align*}
    DCG(rankings) =\sum_{rank} \frac{1}{log_{2}(rank+1)}
\end{align*}
DCG is useful for modeling the popularities of each conspiracy theory group because (1) we wish to model movement between rankings near the top of the Amazon Alexa list as more important than movement near the bottom (\textit{i.e.}, moving from a rank of 100,000 to 10 is more significant than moving from a rank of 900,000 to 800,0010) and (2) DCG accounts for all the rankings of each website in each conspiracy theory list. This is opposed to the median or mean, which do not have both these desired properties. We further note that to prevent us from comparing intra-conspiracy DCGs with different numbers of rankings (\textit{i.e.}, a given website can go in and out of the Alexa rankings without it meaning much for its actual popularity), as recommended~\cite{jarvelin2002cumulated}, we enforce a small minimum value (\textit{i.e.}, rank of 1,000,000$+$) for websites without ranks (we find that this adjustment did not change our results). As previously mentioned, Amazon Alexa changed its methodology for calculating rankings on January 30, 2018~\cite{scheitle2018long, Tajalizadehkhoob2019}. Given this change in methodology, for this analysis, we only utilize rankings and data from February 1, 2018, onward. We further note that we only analyze COVID conspiracy theory website popularities starting from January 1, 2020, onward, given that conspiracy theory websites about the COVID-19 pandemic did not start appearing until then.

For the frequency of hyperlinks to conspiracy theory websites, we utilize the total daily number of hyperlinks from all misinformation websites in our dataset to each category of conspiracy theory websites. To ascertain the date when each hyperlink was published we utilize the Python package \texttt{htmldate}~\cite{barbaresi2020htmldate}. As previously noted, due to the change in the methodology of Amazon Alexa ranking, we only measure the effect of hyperlinks on the popularity of websites from February 1, 2018, onward, filtering out hyperlinks published before then (again making an exception for COVID).

Given the paucity of links from authentic news websites to conspiracy theory websites, we rely on the daily number of authentic news articles mentioning each conspiracy theory topic ("QAnon", "COVID", "Flat-Earth", "UFO", "9/11") as a proxy for story frequency. For example, we count the daily number of authentic news articles that mention ``QAnon'' and measure its effect on the popularity of QAnon websites. In our scrapes of authentic news websites, there are 1.08M pages mentioning UFOs, 29K mentioning Flat-Earth, 1.23M mentioning 9/11, 134K mentioning QAnon, and 8.23M mentioning COVID.\footnote{In the scrapes of our misinformation websites, there are 1.59M pages mentioning UFOs, 108K mentioning Flat-Earth, 1.30M mentioning 9/11, 205K mentioning QAnon, and 3.03M mentioning COVID.} How often authentic news sites have written about each of these conspiracy topics over time is shown in Figure~\ref{fig:conspiracy-num-stories}. We again utilize the Python package \texttt{htmldate} to extract the publication date for each misinformation and authentic news page that we scrape~\cite{barbaresi2020htmldate}. Again, due to the change in the methodology of Amazon Alexa ranking, we only measure the effect of stories on the popularity of conspiracy theory websites from February 1, 2018, onward, (again making an exception for COVID).

We finally note that after fitting our models, to gain a sense of the directionality of the Granger-causal relationships (\textit{i.e.}, if an increase in hyperlinks from misinformation websites to conspiracy theory websites leads to an increase or decrease in the popularity of these conspiracy theory websites), we examine the coefficients of our fit VAR models~\cite{field2018framing}. Specifically, if the fit coefficients of the causal variable are positive on average, then we consider the relationship positive (otherwise negative). This is such that if our model specifies, for example, that more hyperlinks from misinformation websites to conspiracy theory websites would have led to an increase in popularity in these same conspiracy websites, then we consider that relationship to be positive. Finally, we note that given that we test across our five different conspiracy theories, we utilize Benjamini-Hochberg correction procedure~\cite{benjamini1995controlling} for multiple hypothesis testing with a false discovery rate (FDR) of 0.05 to infer partial Granger causality. For details on the Benjamini-Hochberg procedure, see Appendix~\ref{sec:benaj}.

\begin{figure}
\centering
\includegraphics[width=0.85\columnwidth]{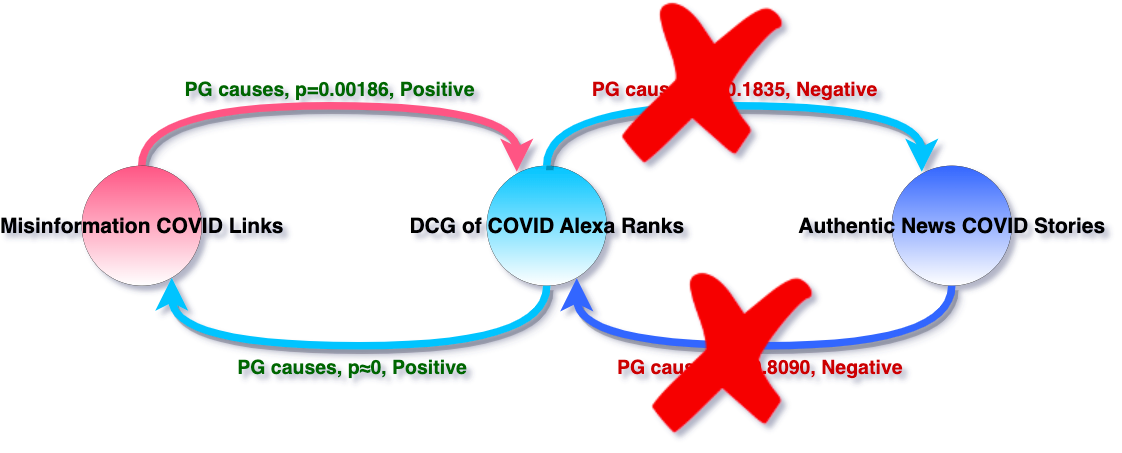} 
\caption{\textbf{COVID Granger causality loops}--- (Non-Granger-causal relationships are Xed) As the number of hyperlinks from misinformation websites to COVID conspiracy theory websites increases this in turn increases the popularity of COVID conspiracy theory websites. As the popularity of these COVID conspiracy theory websites goes up, this in turn increases the number of hyperlinks that misinformation websites utilize. The behavior of authentic news websites appears to not have any correlational effect on the popularity of COVID conspiracy theory websites. }
\label{figure:covid-granger}
\end{figure}

\begin{figure}
\centering
\includegraphics[width=0.88\columnwidth]{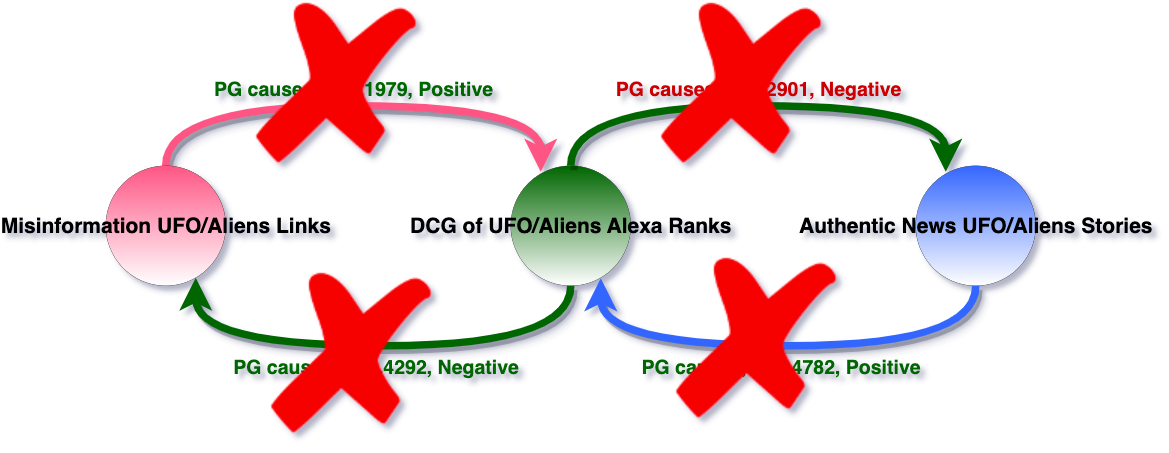} 
\caption{\textbf{UFO/Aliens Granger causality loops}--- (Non-Granger-causal relationships are Xed) There are no statistically significant relationships between the popularity of UFO conspiracy theory websites, hyperlinks from misinformation websites, and the number of stories about UFO/ALiens on authentic news websites.}
\label{figure:ufo-granger}
\vspace{-10pt}
\end{figure}

\begin{figure}
\centering
\includegraphics[width=0.9\columnwidth]{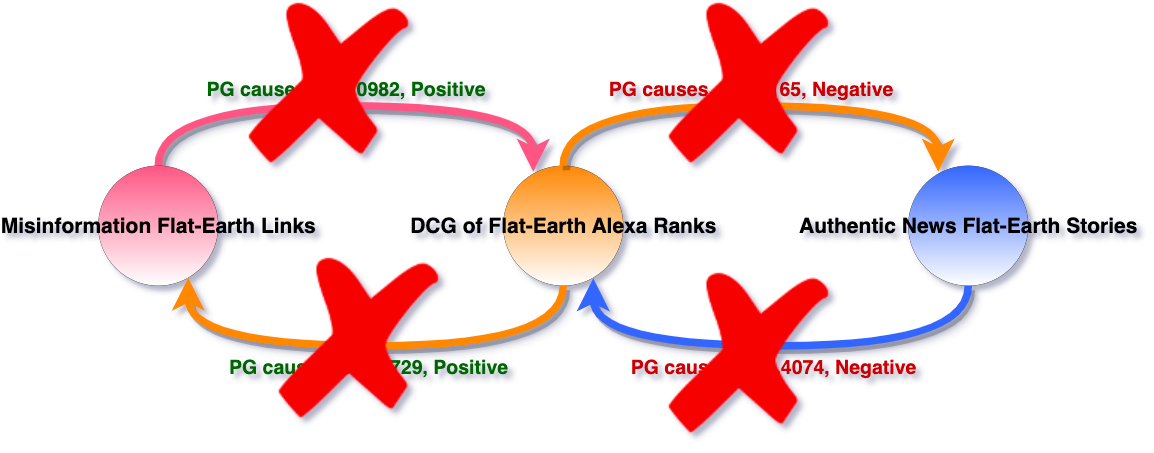} 
\caption{\textbf{Flat-Earth Granger causality loops}--- (Non-Granger-causal relationships are Xed) There are no statistically significant relationships between the popularity of UFO conspiracy theory websites, hyperlinks from misinformation websites, and the number of stories about UFO/Aliens on authentic news websites.}
\label{figure:flat-granger}
\vspace{-10pt}
\end{figure}

\begin{figure}
\centering
\includegraphics[width=0.88\columnwidth]{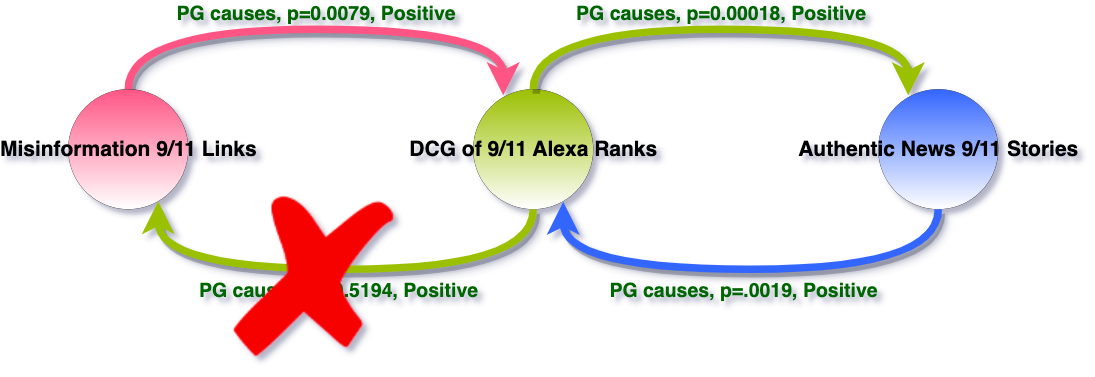} 
\caption{\textbf{9/11 Granger causality loops}--- Non-Granger-causal relationships are Xed) As the number of hyperlinks from misinformation websites to 9/11 conspiracy theory websites increases this, in turn, increases the popularity of 9/11 conspiracy theory websites. As the popularity of these 9/11 conspiracy theory websites goes up, the number of stories that authentic news stories write about 9/11 increases; and finally when authentic news websites write about 9/11 the popularity of 9/11 conspiracy theory websites increases. }
\label{figure:nine11-granger}
\vspace{-10pt}
\end{figure}

\textit{\textbf{Misinformation Websites and Propping up Conspiracy Theories:}}
\begin{figure}
\centering
\includegraphics[width=.85\columnwidth]{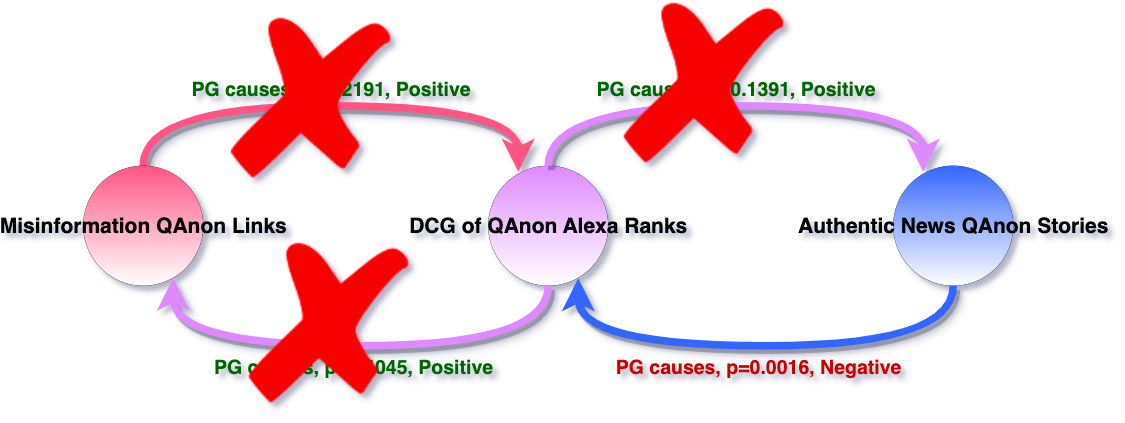} 
\caption{\textbf{QAnon Granger causality loops}--- (Non-Granger-causal relationships are Xed)  As authentic news sites write about QAnon more, the overall popularity of QAnon conspiracy tends to decrease.}
\label{figure:qanon-granger}
\vspace{-10pt}
\end{figure}
In Figures~\ref{figure:qanon-granger}--\ref{figure:nine11-granger}, we see that for the 9/11 and COVID conspiracy theories, the DCG of the popularity of each conspiracy theory group had a positive partial Granger causality with the number of misinformation hyperlinks to these same websites. Simply put, the popularity of conspiracy theory websites correlates positively with the number of hyperlinks from the misinformation outlets to these websites. We thus see, that for these two conspiracy theories, misinformation websites play a role in helping popularize the conspiracy theories. We further observe, for COVID conspiracy theory websites that as these websites became more popular misinformation, outlets wrote about them more, creating a positive feedback loop. Across our remaining conspiracy theories, we observe no statistically significant relationship between the activity of misinformation outlets and the popularity of the group of conspiracy theory outlets. We suspect that across these conspiracy theories other factors (\textit{i.e.}, as social media or societal changes), had a larger and more dominant effect on the changes in the popularity of the conspiracy theories~\cite{hanley2021no,papasavva2021qoincidence}. Altogether, observing positive correlations across all of our conspiracy theories, these results suggest that misinformation websites' behavior corresponds with the popularization and spread of conspiracy theories online, particularly for COVID and 9/11.

\textbf{\textit{Authentic News Websites and Complex Relationships:}} In contrast to hyperlinks from misinformation websites, when authentic news stories do affect the popularity of conspiracy theory websites, the effect appears more complex. For instance, we observe a negative Granger-causal relationship between the popularity of QAnon websites and the number of authentic news QAnon stories (Figure~\ref{fig:conspiracy-num-stories}). For example, we see that the more authentic news websites wrote about QAnon, the more the popularity of QAnon websites decreased. We suspect that as prominent outlets began to write about the conspiracy theory, this led to heightened pressure on platforms to reign in QAnon's spread, decreasing the conspiracy theory's popularity. For example, Reddit~\cite{Tiffany2020}, Twitter~\cite{Timber2020}, and Facebook~\cite{Gatewood2020} all cracked down on the QAnon conspiracy theory following intense media scrutiny of its spread on their platforms~\cite{Sen2020,Greenspan2021,hanley2021no}. In contrast, we observe the opposite relationship for 9/11 websites, observing a positive Granger-casual feedback relationship between the popularity of 9/11 conspiracy theory websites and the number of stories that authentic news publish (Figure~\ref{figure:nine11-granger}). We suspect that as 9/11 annually returns as a topic of interest in the news, 9/11 conspiracy theory websites receive more traffic, and authentic news websites publish more about the event.

Across our remaining conspiracy theories, we do not observe a statistically significant relationship between the number of stories that authentic news websites publish and the popularity of these conspiracy theories. Furthermore, across these conspiracy theories (except UFO/Aliens), we observe a negative correlation between the number of stories that these authentic news outlets published and the popularity of the corresponding conspiracy theory group. This suggests, that while reporters often worry about accidentally redirecting readers or exposing their audience to conspiracy theories from writing about them~\cite{tsfati2020causes}, we see largely little to mixed (\textit{e.g.}, 9/11) evidence for this phenomenon.

We finally note that while correlational, these results show that both misinformation and authentic news activity \textit{do} appear to correspond with the popularity of different conspiracy theories \textit{and} in different ways depending on the conspiracy theory.

\section{Discussion}
In this work, we utilize readily identifiable conspiracy theory website groups to understand how conspiracy theories have influenced and affected the Internet, particularly the news media ecosystem over the past decade. From our initial analysis, we see that conspiracy theory websites share a similar ecosystem with misinformation websites. We further observe instances of self-reinforcing relationships between misinformation and conspiracy theory websites (\textit{e.g.}, COVID), finding positive partial Granger-causal loops.  Now, in this section, we give an overview of the implications of these results.

\subsection{The Interdependence Among Conspiracy Theories and Reliance on News Media} As seen throughout this work, and largely as expected, conspiracy theory websites often hyperlink to and utilize the same domains. While each group of conspiracy theory websites in our dataset was distinct, we find that every group points to at least 7.5\% of the same domains as every other group, with QAnon even connecting to 35.6\% of the same domains as COVID websites (Table~\ref{table:conspiracy-domain-connections}). This reinforces prior research that individuals who believe in one conspiracy theory are more likely to believe in another~\cite{goertzel1994belief}. This is in contrast to the misinformation, authentic news,  and non-news domain which share at most 4.8\% (QAnon) with any of our conspiracy theory website groups and as little as 0.55\% (COVID) (Figure~\ref{figure:conspiracies-to-different-types}).

Despite misinformation, authentic news, and non-news websites having substantially fewer connections to the same websites that conspiracy theory websites connect to, we do see that conspiracy theory websites have extensively utilized the same sources and domains as authentic news and misinformation outlets. Conspiracy theories websites furthermore heavily cited many of the specific news outlets in our lists of misinformation and authentic news outlets, often referencing nytimes.com, cnn.com, foxnews.com, as well as  zerhohedge.com, globalresarch.ca, and breitbart.com. This illustrates the degree to which conspiracy theories, despite being fringe, actually heavily rely on popular outlets, exposing their adherents to ``mainstream'' news. 

\subsection{Conspiracy Theories on the Internet}
We unsurprisingly find that QAnon is one of the largest and most prolific conspiracy theories on the Internet. Given QAnon has become a repository for many different conspiratorial beliefs, we argue that understanding exactly how this conspiracy theory works and de-radicalizing its adherents can help de-radicalize those influenced by other conspiracy theories.

While some have argued that conspiratorial thinking has remained constant throughout human history, we also see that several conspiracy theories have become a larger and larger part of the hyperlinks on misinformation websites during the past decade. We indeed see that since 2008, the percentage of conspiracy-oriented links by misinformation websites has increased substantially. Besides misinformation websites, however, we find that other types of websites have largely not promoted conspiracy theory websites or conspiratorial material. This suggests somewhat of a divergence of the types of websites online: those prone to conspiracy theories and those that are not. While it is uncertain whether the number of true believers of conspiracies has increased since 2008, misinformation websites' audiences \textit{have} been exposed to more conspiratorial content.

\subsection{The Role of Misinformation in Promoting Conspiracy Theories}
We find that misinformation websites' role goes further than merely hyperlinking to conspiracy theory websites. As misinformation outlets publish more hyperlinks to conspiracy theory websites this, in turn, correlates with the increased popularity of these websites for both the COVID and 9/11 conspiracy theories. Conversely, we observe some evidence (9/11 and QAnon) that authentic news websites writing about conspiracy theories can potentially have an effect on the popularity of these conspiracy theories, but often in complex ways. If platforms and researchers want to tackle conspiracy theories and their potentially detrimental effects, we argue that we should focus our efforts on curtailing the role of large misinformation outlets whose activity often positively correlates with the popularity of particular conspiracy theories. Given the uptick in exposure to conspiracy theories from misinformation websites, online platforms must take action against the most egregious actors. 


\section{Conclusion}
In this work, we studied the relationships between conspiracy theories, misinformation, and political polarization on the Internet by detailing the inner workings and interactions of 755~websites dedicated to conspiracy theories and prominent news outlets. We found that QAnon remains the most prominent conspiracy theory on the Internet, feeding people into other online conspiratorial movements. We then detailed how misinformation websites in particular connect people to conspiracy materials in contrast to authentic news websites and non-news websites. Using the notion of partial Granger causality, we detailed the correlative relationships that exist between the number of hyperlinks on misinformation websites and the popularity of these conspiracy theory websites, also describing the more complex relationship that authentic news websites have with conspiracy theory websites. 


\section*{Acknowledgements} 
This work was supported in part by the National Science Foundation under grant \#2030859 to the Computing Research Association for the CIFellows Project, a gift from Google, Inc., NSF Graduate Fellowship DGE-1656518, and a Stanford Graduate Fellowship.

\bibliographystyle{ACM-Reference-Format}
\bibliography{sample-base}

\appendix
\section{Granger Causaility}\label{sec:granger-causality}
Granger causality is a means of measuring if one time series is useful for forecasting another. Granger causality compares the efficacy of predicting a stochastic process $Y$ using \textit{all the information in the universe} $U$ with the efficacy of using all information except for some other stochastic process $X$, denoted $U\backslash X$. If removing $X$ reduces the predictive power of $U$ regarding $Y$, then $X$ Granger causes $Y$. More formally:
\begin{enumerate}
    \item Let $X$ and $Y$ be stationary stochastic processes.
    \item Denote $U_i$ as all the information in the universe $U$ up to time $i$ ($U_{i-1},..., U_{i-\infty}$), $Y_i$ as all the information in $Y$ up to time $i$ ($Y_{i-1},..., Y_{i-\infty}$), and $X_i$ as all the information in $X$ up to time $i$ ($X_{i-1},..., X_{i-\infty}$).
    \item Denote $\sigma^2 (Y_i|U_i$) as the variance of the residual from the prediction of $Y_i$ using $U_i$ at time $i$.
    \item Denote $\sigma^2 (Y_i|U_i \backslash X_i$) as the variance of the residual from the prediction using all the information in $U_i$ at time $i$ except for $X_i$.
\end{enumerate}

\noindent
if $\sigma^2 (Y_i|U_i) < \sigma^2 (Y_i|U_i\backslash X_i)$ then $X$ Granger-causes $Y$ or $X \rightarrow Y$. If $Y$ also  Granger-causes $X$ then we also say, $X \leftrightarrow Y$ and \textit{feedback} is occurring~\cite{granger1969investigating}. 

\section{Partial Granger Causality}\label{sec:p-granger-causality}
Previous works have found that repeatedly applying the original notions of Granger causality to identify causal relationships can lead to misleading results~\cite{chen2004analyzing,guo2008partial}. An extended form of Granger causality, conditional Granger (CG) causality attempts to correct for the interaction of multiple times series while identifying relationships. CG causality thus determines whether one time-series Granger causes a second time-series conditional on a third or fourth time series. However, CG causality assumes that all relevant variables have been included in the model. Given that we cannot fully model \textit{all} interactions amongst news, misinformation, and conspiracy theory websites, we instead utilize an extended form of CG causality namely \textit{partial Granger Causality} to determine and understand these interactions.

Time-domain partial Granger Causality specifically models exogenous environmental variables and unmeasured endogenous variables as well as conditioned interactions to mitigate confounding effects~\cite{guo2008partial,yurdakul2015determinants}. Specifically, to determine the Granger-causal impact of a time-series $Y_t$ on another time series $X_t$ given the influence of a third time-series $Z_t$ as well as exogenous inputs and unmeasured latent models, PGC models the three time series together in a restricted and an unrestricted set of linear vector autoregressive (VAR) models. The restricted model assumes that the two time series $X_t$ and $Z_t$ are linearly dependent on past values of themselves, as well as time-dependent exogenous inputs, endogenous latent variables, and noise terms. 

\begin{align*}
X_t &= \Sigma^p_{i=1}a_{1i}X_{t-i} + \Sigma^p_{i=1}b_{1i}Z_{t-i} +\overrightarrow{\epsilon_{1t}} +\overrightarrow{\epsilon_{1t}^E}+\overrightarrow{B_1(L)\epsilon_{1t}^L}\\
Z_t &= \Sigma^p_{i=1}c_{1i}X_{t-i} + \Sigma^p_{i=1}d_{1i}Z_{t-i} +\overrightarrow{\epsilon_{2t}} +\overrightarrow{\epsilon_{2t}^E}+\overrightarrow{B_2(L)\epsilon_{2t}^L}
\end{align*}
with noise covariance matrix $S$ and noise terms $e_{it}$,
\begin{align*}
S &= \begin{bmatrix} var(e_{1t}), cov(e_{1t},e_{2t}) \\
 cov(e_{2t},e_{2t}), var(e_{2t})
\end{bmatrix} \\
e_{it} &= \overrightarrow{\epsilon_{it}} +\overrightarrow{\epsilon_{it}^E}+\overrightarrow{B_i(L)\epsilon_{it}^L}
\end{align*}
where p is lag/the number of past terms that affect the current values of the time-series, matrices  $A_1$, $B_1$, $C_1$, and $D_1$ contain coefficients that model these autogressive effects,   $\overrightarrow{\epsilon_{i}}$ model Gaussian white noise processes, $\overrightarrow{\epsilon_{i}^E}$ model zero-mean exogenous Gaussian variables, and $B_1(L)\overrightarrow{\epsilon_{i}^L}$ model latent variables. Specifically $B_1(L)\overrightarrow{\epsilon_{it}^L}$ incorporate latent variables by assuming that the $t$th network element receives unmeasured inputs of the form $\Sigma^N_jx_{tj}/N $ where here each $x_{tj}$ is a stationary time series and $j$ is the latent index. 

In contrast, the unrestricted model assumes  $X_t$ and $Z_t$, as well as $Y_t$, are all linearly dependent on the past values of these three time series as well as time-dependent exogenous inputs, endogenous latent variables, and noise terms.

\begin{align*}
X_t &= \Sigma^p_{i=1}a_{2i}X_{t-i} +\Sigma^p_{i=1}b_{2i}Y_{t-i}+ \Sigma^p_{i=1}c_{2i}Z_{t-i} +\overrightarrow{\epsilon_{3t}} +\overrightarrow{\epsilon_{3t}^E}+\overrightarrow{B_3(L)\epsilon_{3t}^L}\\
Y_t &= \Sigma^p_{i=1}d_{2i}X_{t-i} +\Sigma^p_{i=1}f_{2i}Y_{t-i}+ \Sigma^p_{i=1}g_{2i}Z_{t-i} +\overrightarrow{\epsilon_{4t}} +\overrightarrow{\epsilon_{4t}^E}+\overrightarrow{B_4(L)\epsilon_{4t}^L} \\
Z_t &= \Sigma^p_{i=1}h_{2i}X_{t-i} +\Sigma^p_{i=1}k_{2i}Y_{t-i}+ \Sigma^p_{i=1}m_{2i}Z_{t-i} +\overrightarrow{\epsilon_{5t}} +\overrightarrow{\epsilon_{5t}^E}+\overrightarrow{B_5(L)\epsilon_{5t}^L}
\end{align*}
with noise covariance matrix $\Sigma$ and noise terms $e_{it}$,
\begin{align*}
\Sigma &= \begin{bmatrix} var(e_{3t}), cov(e_{3t},e_{4t}), cov(e_{3t},e_{5t})  \\
cov(e_{4t},e_{3t}), var(e_{4t}), cov(e_{4t},e_{5t})  \\
cov(e_{5t},e_{3t}), cov(e_{5t},e_{4t}),var(e_{5t})  \\
\end{bmatrix} \\
e_{it} &= \overrightarrow{\epsilon_{it}} +\overrightarrow{\epsilon_{it}^E}+\overrightarrow{B_i(L)\epsilon_{it}^L}
\end{align*} Taking this all into account, partial Granger causality calculates the ratio of (1) a measure based on the noise covariance matrix $S$ in the first model of the accuracy of predicting the present value of X using the history of X, conditioned on Z and eliminating the influence of the exogenous variables $\overrightarrow{\epsilon_{i}^E}$ and the latent variables $\overrightarrow{B_i(L)\epsilon_{i}^L}$, and (2) a measure based on the noise covariance matrix $\Sigma$ in the second model of the accuracy of predicting the present value of X using on the history of both X \textit{and Y} conditioned on Z and eliminating the influence of the exogenous variables $\overrightarrow{\epsilon_{i}^E}$.  
 
\begin{align*}
    F_1 = ln \left( \frac{S_{11}-S_{12}^{-1}S_{21}}{\Sigma_{11}-\Sigma_{12}^{-1}\Sigma_{21}} \right)
\end{align*}
As specified~\cite{guo2008partial,roelstraete2012does} the distribution of $F_1$ can be used to determine whether $Y_t$ Granger-causes $X_t$. Using the same noise covariance matrices, whether $Y_t$ Granger-causes $X_t$ can be determined as well as $Z_t$ Granger-causing $Y_t$, \textit{etc...} Thus to test for Granger causality between two time-series $X_t$ and $Y_t$ conditioned on a third time-series $Z_t$ and taking into account exogenous environmental variables as well as endogenous latent variables, we test against the null hypothesis that the second model does not add information in predicting future values of $X_t$.


\section{Bejamini-Hochberg Procedure}\label{sec:benaj}

The Benjamini-Hochberg procedure is means of controlling the false discovery rate when performing multiple statistical tests. The procedure, after performing all tests and acquiring their individual p-values is as follows:
\begin{enumerate}
\item Sort the p-values in ascending order.
\item Assign ranks to the p-values of each test. 
\item Calculate each individual p-value’s Benjamini-Hochberg critical value, using the formula:
\begin{align*}
 (i/m)Q
\end{align*} 
where $i$ is the individual p-value’s rank, $m$ is the total number of tests, and $Q$ is the false discovery rate (FDR).
\item Compare the original p-values to the critical Benjamini-Hochberg values from Step 3. If the original p-value is smaller than its critical values, then reject the null hypothesis for that given statistical test; otherwise, accept the null hypothesis. 
\end{enumerate}

\end{document}